\documentclass[
 reprint,
 amsmath,amssymb,
 aps,
 superscriptaddress, 
 floatfix, 
]{revtex4-2}

\usepackage{dcolumn}
\usepackage{bm}
\usepackage[pdftex]{graphicx,color} 
\usepackage{comment}
\usepackage{natbib}
\usepackage{breakcites}

\usepackage{here}
\usepackage[normalem]{ulem}
\usepackage[breaklinks=true]{hyperref}
\usepackage{braket}
\usepackage{url}
\usepackage{multirow}

\usepackage[whole]{bxcjkjatype} 

\begin{document}

\title{
Inverse Hamiltonian design of highly-entangled quantum systems 
}

\author{Koji Inui}
\affiliation{Department of Applied Physics, The University of Tokyo, Hongo, Tokyo 113-8656, Japan}
\affiliation{Department of Nuclear Engineering and Management, The University of Tokyo, Hongo, Tokyo 113-8656, Japan}
\author{Yukitoshi Motome}
\affiliation{Department of Applied Physics, The University of Tokyo, Hongo, Tokyo 113-8656, Japan}

\begin{abstract}
Solving inverse problems to identify Hamiltonians with desired properties holds promise for the discovery of fundamental principles. 
In quantum systems, quantum entanglement plays a pivotal role in not only characterizing the quantum nature but also developing quantum technology like quantum computing. 
Nonetheless, the design principles of the quantum entanglement are yet to be clarified. 
Here we apply an inverse design framework using automatic differentiation to quantum spin systems, aiming to construct Hamiltonians with large quantum entanglement. 
We show that the method automatically finds the Kitaev model with bond-dependent anisotropic interactions, whose ground state is a quantum spin liquid, on both honeycomb and square-octagon lattices. 
On triangular and maple-leaf lattices with geometrical frustration, it generates numerous solutions with spatially inhomogeneous interactions rather than converging to a specific model, but it still helps to construct unprecedented models. 
The comparative study reveals that bond-dependent anisotropic interactions, rather than isotropic Heisenberg interactions, amplify quantum entanglement, even in systems with geometrical frustration.
The present study paves the way for the automatic design of new quantum systems with desired quantum nature and functionality. 
\end{abstract}

\maketitle

\section{Introduction}
\label{introduction}


Quantum entanglement holds pivotal significance across a wide range of research fields, from black holes in cosmology~\cite{RevModPhys.93.035002} to quantum materials in condensed matter physics~\cite{RevModPhys.80.517}. 
Given that entanglement represents nonlocal correlations, systems with pronounced quantum entanglement can manifest properties that are not accessible to classical descriptions, potentially paving the way for advances in quantum information processing. 
For instance, the efficiency of quantum error correction is intricately tied to quantum entanglement measures~\cite{PhysRevA.54.3824,PhysRevLett.97.150504,PhysRevLett.102.190501,PhysRevLett.125.030505,PhysRevB.103.104306}. 
The topological entanglement entropy, one of such entanglement measures, in stabilizer codes has been intensively studied~\cite{PhysRevB.76.184442,PhysRevB.97.125101,PhysRevB.84.195120}.
In condensed matter physics, quantum phase transitions can be detected by the entanglement measures~\cite{PhysRevLett.90.227902,PhysRevA.66.032110,Osterloh2002,PhysRevA.66.032110,PhysRevB.96.054503,PhysRevB.78.224413,PhysRevB.89.104303,PhysRevD.93.126004,PhysRevLett.119.225301,RevModPhys.80.517}, 
as is the classification of topological orders~\cite{PhysRevLett.96.110404,PhysRevLett.96.110405,Jiang2012,Isakov2011}. 
Furthermore, the entanglement entropy (EE) serves to characterize black holes~\cite{Solodukhin2011,Hartman2013} and the many-body localization~\cite{Bauer_2013}. 

Much of the existing research has focused on assessing quantum entanglement measures of specific quantum systems. 
Yet, for applications like the synthesis of quantum materials with unique properties and the development of devices suitable for quantum computation, the primary requirement is designing systems with the desired entanglement properties. 
Nevertheless, there remains a limited understanding of the design principles of quantum entanglement.

An efficient way to obtain the systems that exhibit the desired properties is to solve the inverse problem. 
Thus far, numerous methods have been proposed for such inverse design. 
In particular, for quantum systems, several techniques were developed, including the quantum state preparation~\cite{PhysRevA.95.042318} and the parent Hamiltonian identification with specific wave functions~\cite{PhysRevA.79.032504,PhysRevX.8.031029,Pakrouski2020automaticdesignof,Qi2019determininglocal,PhysRevB.98.081113,PhysRevB.97.075114,tang2023nonhermitian}. 
In these methods, however, it is not straightforward to directly derive systems from tangible physical quantities or entanglement measures. 
Machine learning-driven approaches, such as the Bayesian optimization~\cite{PhysRevB.95.064407}, the generative models~\cite{doi:10.1126/science.aat2663}, and the random forests~\cite{PhysRevResearch.3.013132}, have also been explored. 
While these methods are potentially powerful, they require extensive data collections and face challenges in ensuring precision. 
An alternative method using automatic differentiation, which was recently developed by the authors~\cite{Inui2023}, eliminates the need for data and addresses the challenges mentioned above. 
Thus, it can be a versatile tool for designing the quantum nature of the system like quantum entanglement, but it has been applied only to quantum systems that can be dealt with independent particle approximations.

In this study, we develop a method to design Hamiltonians of quantum many-body systems characterized by large quantum entanglement, leveraging the inverse design framework utilizing automatic differentiation~\cite{Inui2023}. 
While the framework itself is versatile, the straightforward application by using, e.g., the EE, does not work efficiently, leading to instability of the optimization. 
To circumvent the difficulty, we develop two techniques: (i) We define and use the thermal ensemble of entanglement entropy (TEEE) as the objective function, and (ii) we incorporate a symmetrization in the optimization procedure such that the result is spatially homogeneous.
It is noteworthy that our method can be extended to other entanglement measures such as logarithmic negativity (LN)~\cite{PhysRevLett.95.090503}, mutual information (MI)~\cite{PhysRevLett.78.2275}, and topological entanglement entropy~\cite{PhysRevLett.96.110404,PhysRevLett.96.110405}. 
In addition, it can be adapted to design systems with predetermined entanglement properties instead of maximizing them.

We apply this method to quantum spin models with two-spin interactions on various lattice structures with and without geometrical frustration. 
For the models on the honeycomb and square-octagon lattices, both of which are free from geometrical frustration, we show that the optimized results are equivalent to the Kitaev model~\cite{KITAEV20062}, a renowned exactly-solvable quantum spin liquid model and a potential candidate for topological quantum computation. 
In contrast, for the models on the triangular and maple leaf lattices with geometrical frustration, the optimization does not settle on a specific model, and instead, it yields numerous solutions with spatially inhomogeneous interactions, which show similar quantum entanglement. 
Even in these cases, however, we demonstrate that the present method is useful to find a homogeneous Hamiltonian with large entanglement; as an example, we obtain a model only with anisotropic interactions on the maple-leaf lattice. 
This underscores the proficiency of our method in autonomously generating models with large quantum entanglement.

Furthermore, we investigate the characteristics of interactions that are dominant in the obtained models with large entanglement. 
Our findings emphasize that systems with bond-dependent anisotropic interactions like the Kitaev-type interaction are more inclined to have enhanced entanglement, irrespective of the presence or absence of geometrical frustration. 
For instance, on the triangular lattice, large entanglement is achieved not by isotropic Heisenberg interactions expected for the resonating valence bond state~\cite{ANDERSON1973153}, but by strongly anisotropic interactions depending on the bonds. 
This sheds light on the intricate relationship between quantum entanglement and the form of exchange interactions. 
Advancing and refining this method could pave the way for the discovery of systems that are pivotal for cutting-edge applications, such as quantum computing.

This paper is organized as follows. 
In Sec.~\ref{sec:framework}, we introduce the method. 
After introducing the general framework, we propose the objective function in Sec.~\ref{subsec:objectivefunction} and describe the actual calculation flow in Sec.~\ref{subsec:calculationflow}. 
In Sec.~\ref{sec:result}, we present the results for quantum spin systems on honeycomb (Sec.~\ref{subsec:honeycomb}), square-octagon (Sec.~\ref{subsec:48}), triangular (Sec.~\ref{subsec:tri}), and maple-leaf lattices (Sec.~\ref{subsec:maple}). 
We discuss the results for different lattices from the view point of the ratio of anisotropic interactions in Sec.~\ref{sec:discussion}. 
Finally, we give the concluding remarks in Sec.~\ref{sec:conclusion}.

\section{Method}
\label{sec:framework}

In this section, we introduce a method to obtain the model with large quantum entanglement, based on the framework of inverse Hamiltonian design using automatic differentiation~\cite{Inui2023}. 
In this framework, the Hamiltonian is parameterized by a number of parameters \( \bm{\theta} \) as \( \mathcal{H}(\bm{\theta}) \). 
We also construct an objective function, \( L(\bm{\theta}) \), specifically designed such that its minimum corresponds to the realization of a desired state.
Utilizing the gradient \( \partial L(\bm{\theta})/\partial \bm{\theta} \),  efficiently computed through automatic differentiation, we iteratively update \( \bm{\theta} \) using the gradient descent method to minimize \( L(\bm{\theta}) \).
This facilitates the design of a Hamiltonian having the desired properties. 
In this study, we apply the method to the design of quantum many-body systems with large quantum entanglement. 
In the following subsections, we detail the objective function and provide the comprehensive calculation flow.

\subsection{Objective function}
\label{subsec:objectivefunction}

A prevalent measure of quantum entanglement is the EE, derived from the density matrix, \( \rho = |\psi\rangle \langle \psi | \), where \( |\psi\rangle \) represents the wave function of the systems~\cite{PhysRevA.53.2046}. 
The EE between bi-partitioned subsystems A and B is given by
\begin{align}
S_{\rm A} = - \rm{Tr}_{\rm A} \rho_{\rm A} \log \rho_{\rm A}, \label{eq:S} 
\end{align}
with $\rho_{\rm A} = \rm{Tr}_{\rm B} \rho $. 
Here, \( \rm{Tr}_{A(B)} \) indicates the partial trace over the subsystem A (B). 
We find, however, that straightforward application of the EE to the objective function in the framework of the inverse Hamiltonian design leads to instability in the optimization process. 
This is because the energy levels of the ground state and the first excited state are swapped during the optimization process, resulting in abrupt changes in the value of the objective function. 
Another problem is that the value of EE depends on how the system is bi-partitioned.


To address these issues, we define the TEEE, which aggregates the EE with the Boltzmann distribution, as
\begin{align}
S^T_{g, \xi}(\bm{\theta}) = \frac{\sum_n S_{n,g,\xi}(\bm{\theta}) \exp \bigl(-\beta E_n(\bm{\theta})  \bigr)}{\sum_n \exp \bigl(-\beta E_n(\bm{\theta})\bigr)}, 
\label{eq:SCg} 
\end{align}
where $\beta$ is the inverse temperature, and \( E_n({\bm{\theta}}) \) and \( S_{n,g,\xi}({\bm{\theta}}) \) denote the energy and EE for each eigenstate \( n \), respectively; 
\( g \) represents the groups of bi-partitioning patterns associated with translational and rotational symmetries of the system, and \( \xi \) represents the bi-partition index within a group \( g \).
For instance, for a cluster with four sites $ \{s_1,s_2,s_3,s_4\}$ in a periodic one-dimensional chain, there are two groups of bi-partitioning patterns, one with two elements and the other with one element, represented as $(g=1, \xi=1) = (s_1,s_2 | s_3,s_4) $ and $(g=1,\xi=2) = (s_2,s_3|s_1,s_4)$, and $(g=2,\xi=1)= (s_1,s_3|s_2,s_4) $. 
Employing the TEEE mitigates the computational instability mentioned above, as it incorporates not merely the ground state but also the excited states. 
In addition, to resolve the problem that the EE depends on the bi-partitioning pattern, we construct the objective function $L$ including a penalty to suppress the deviation of TEEE between different $\xi$ within each group $g$ as 
\begin{align}
L(\bm{\theta}) &= -\bar{S}^T(\bm{\theta}) + \lambda \Delta S^T(\bm{\theta}) \label{eq:L}, 
\end{align}
where 
\begin{align}
\bar{S}^T(\bm{\theta}) &= \frac{1}{\sum_g N_{g}} \sum_{g\xi} S^T_{g,\xi}(\bm{\theta}) \label{eq:barS}, \\
\Delta S^T(\bm{\theta}) &= \sum_g \sqrt{\frac{1}{N_g} \sum_\xi (S^T_{g,\xi}(\bm{\theta}) - \bar{S}^T_g(\bm{\theta}))^2 },  \label{eq:stdS}
\end{align}
with
\begin{align}
\bar{S}^T_g(\bm{\theta}) &= \frac{1}{N_g}\sum_\xi  S^T_{g,\xi}(\bm{\theta}).  \label{eq:barSg}
\end{align}
Here, \( N_g \) denotes the number of bi-partitioning patterns within each group \( g \), and \( \lambda \geq 0 \) is a hyperparameter for the penalty term. 
The first term in Eq.~(\ref{eq:L}) aims to enhance the overall TEEE, while the second term ensures uniformity of TEEE within the group. 
We also tried other objective functions to increase the MI or the LN instead of the TEEE, and found that the TEEE is most suitable for the present purpose;
this is because it allows solutions with degeneracy between states connected by symmetry, whereas the MI and LN only allow nondegenerate solutions (see Appendix~\ref{sec:mutinfo}). 

\subsection{Calculation flow and computational details}
\label{subsec:calculationflow}

\begin{figure}[htbp]
        \centering
        \includegraphics[width=1.0\columnwidth,pagebox=cropbox,trim={0 0 0 0},clip]{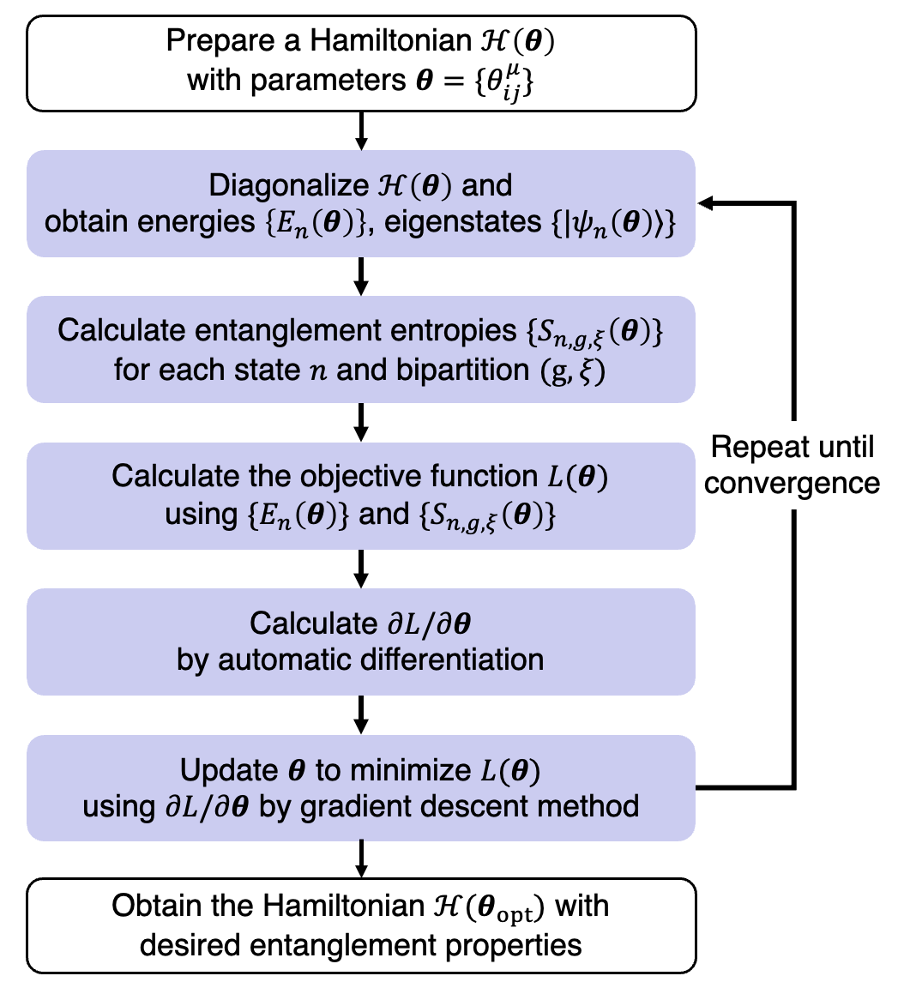} 
        \caption{
        Flowchart of the inverse design of Hamiltonians with desired quantum entanglement properties using automatic differentiation. 
        See the main text for details. 
}
        \label{Fig:procedure}
\end{figure}

The overall calculation procedure is illustrated in Fig.~\ref{Fig:procedure}. 
First, we define a Hamiltonian depending on parameters \(\bm{\theta}\). 
Next, the Hamiltonian is diagonalized, yielding the energy \(E_n(\bm{\theta})\) and the wave function \(|\psi_n(\bm{\theta})\rangle\) for each eigenstate \(n\). 
Then, we compute the density matrix as \(\rho_n(\bm{\theta}) = |\psi_n(\bm{\theta})\rangle \langle \psi_n(\bm{\theta})|\) for every state $n$. 
Subsequently, the EE \(S_{n,g,\xi}(\bm{\theta})\) is computed for each state with respect to each bi-partitioning pattern \(g,\xi\). 
Through the calculation of \(S^T_{g,\xi}\) in Eq.~(\ref{eq:SCg}), we compute the objective function \(L(\bm{\theta})\) as given in Eq.~(\ref{eq:L}). 
Once \(L(\bm{\theta})\) is obtained, its gradient with respect to \(\bm{\theta}\), \( \partial L(\bm{\theta}) / \partial {\bm{\theta}}\), is computed using automatic differentiation. 
Then, \(\bm{\theta}\) is updated to minimize \(L(\bm{\theta})\) using \( \partial L(\bm{\theta}) / \partial {\bm{\theta}}\) by the gradient descent method. 
Iteratively applying this procedure, we are able to construct a Hamiltonian, $\mathcal{H}(\bm{\theta}_{\rm opt})$, with large and spatially uniform quantum entanglement, where $\bm{\theta}_{\rm opt}$ is parameters after the optimization.

While our approach is applicable to arbitrary Hamiltonians, we focus on a quantum spin Hamiltonian with two-spin exchange interactions between spin-\(\frac12\) moments, which is given by  
\begin{align}
\mathcal{H}(\bm{\theta}) = \sum_{\langle ij \rangle \mu} J_{ij}^{\mu} \hat{\sigma}_i^\mu \hat{\sigma}_j^\mu, \label{eq:Ham} 
\end{align}
where $\mu$ spans $\{x,y,z\}$, \(i,j\) are indices designating individual sites, and \(\hat{\sigma}_i^\mu\) represents the Pauli operator at site \(i\). 
We parametrize the exchange constants by \(\bm{\theta}=\{\theta_{ij}^\mu \}\) as 
\begin{align}
J_{ij}^{\mu} = \frac{\theta_{ij}^\mu}{\sqrt{\sum_\mu {\theta_{ij}^\mu}^2}},  \label{eq:Jij}
\end{align}
which satisfies the condition \(\sum_\mu (J_{ij}^{\mu})^2 = 1\). 
We note that the parametrization is not unique, and the optimized results may depend on the way of parametrization as well as the objective function;
see Appendix~\ref{sec:app_con}.  
While the present study deals with only the diagonal interactions in spin space, it can be straightforwardly extended to include off-diagonal components such as $J_{ij}^{xy}$, as well as other forms of interactions rather than the two-spin exchange interactions for arbitrary spin length $S$. 


We apply the following numerical techniques to make the optimization better and more efficient. 
First, a tiny random hermitian matrix, whose element is in the order of \(10^{-8}\), is added to the Hamiltonian to avoid the instability caused by degeneracy in the energy levels.  
This step is crucial because the current version of the JAX library lacks support for the automatic differentiation of diagonalization when involving degeneracy in eigenvalues. 
The minor modification slightly lifts the degeneracy, thereby stabilizing the optimization process. 
See Appendix~\ref{sec:app_instability} for the details. 
Next, in the calculation of $S^T_{g,\xi}$, only $N$ states from the lowest energy are used out of $2^{N_S}$ total states for the number of spins $N_S$.
It is because high energy states do not contribute to the value of TEEE if $\beta$ is large enough. 
We set \(N=40\) throughout the paper;
increasing $N$ has a negligible effect on the results. 
To compute the EE, we use the relation \({\rm Tr}\rho_{\rm A} \log \rho_{\rm A} = \sum_i \gamma_i \log \gamma_i\), where \(\{ \gamma_i\}\) represents the eigenvalues of \(\rho_{\rm A}\).
For the gradient decent method, we employ AdaBelief optimizer~\cite{NEURIPS2020_d9d4f495} with hyperparameters \(\beta_1=0.9\) and \(\beta_2=0.999\).
The hyperparameter $\lambda$ and the learning rate \(\eta\) are adjusted in accordance with optimization steps.

We conduct calculations with 100 different initial conditions. 
In case the optimized Hamiltonians have spatially inhomogeneous interactions, we may further optimize to purify the Hamiltonian with another objective function, as shown in Appendix~\ref{sec:S_params}.
Automatic differentiation is performed using the JAX library in Python~\cite{jax2018github}, and optimization is carried out with the optax library. 
All computations are executed on an NVIDIA A100 GPU.

\section{Results}
\label{sec:result}

We apply the method to the model in Eq.~(\ref{eq:Ham}) on four different lattices: the honeycomb lattice, the square-octagon lattice, the triangular lattice, and the maple leaf lattice. 
The former two are bipartite, while the latter two include triangles, i.e., geometrically frustrated. 
We present the results on each lattice one by one in the following sections. 

\subsection{Honeycomb lattice}
\label{subsec:honeycomb}

\begin{figure*}[htbp]
        \centering
        \includegraphics[width=2.0\columnwidth,pagebox=cropbox,trim={0 0 0 0},clip]{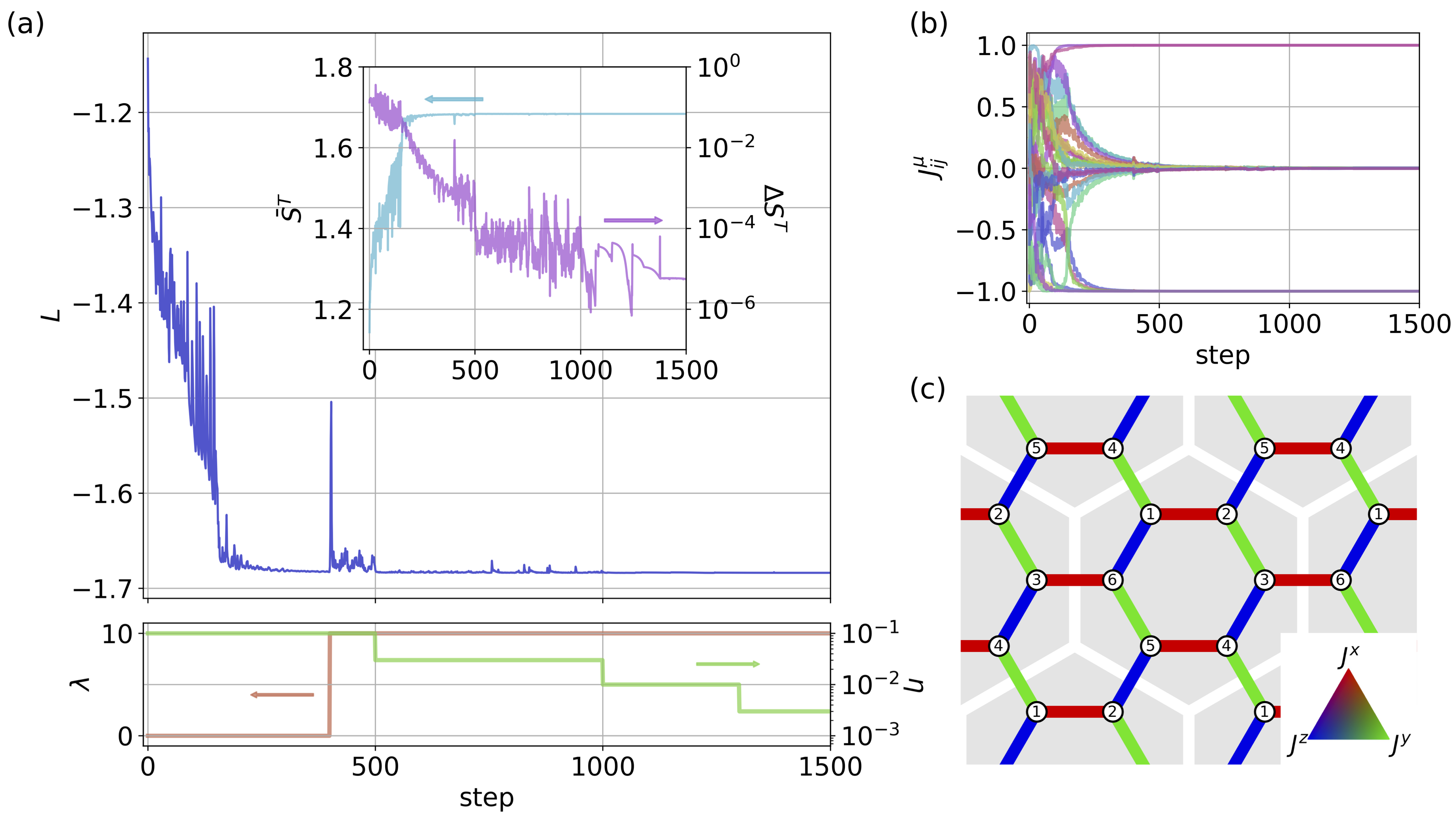} 
        \caption{
        A representative optimization result for the honeycomb lattice model. 
        (a) Changes in the objective function $L$ through the optimization process. The inset shows changes in $\bar{S}^T$ in Eq.~(\ref{eq:barS}) and $\Delta S^T$ in Eq.~(\ref{eq:stdS}). 
        The lower panel shows the schedule of $\lambda$ in Eq.~(\ref{eq:L}) and the learning rate $\eta$. 
        (b) Changes in each component of spin interactions. 
        (c) Optimized interaction parameters. 
        The shades denote six-site cluster unit cells. 
        The color of the bonds represents each spin interaction according to the color code in the inset which shows the ratio of the elements of $J$, e.g., red denotes that $J^x=1$ and $J^y=J^z=0$. 
        While the optimized results in Fig.~\ref{Fig:honeycomb}(b) have a mixture of both positive and negative values, we plot the absolute values in (c) since 
        the value of $L$ remains consistent when all these signs of $J^\mu_{ij}$ are entirely negative. 
}
        \label{Fig:honeycomb}
\end{figure*}

First, we apply the method to the honeycomb lattice consisting of six-site cluster with periodic boundary conditions, illustrated in Fig.~\ref{Fig:honeycomb}(c). 
This lattice comprises nine distinct bonds and 27 parameters. 
We restrict our consideration to bi-partitions where the system is divided into two interconnected subsystems, with each subsystem having three sites, as detailed in Table \ref{table:honeycomb}. 
In this case, we have only one group which contains nine different elements. 
The inverse temperature is maintained at \(\beta=60\) throughout the calculation.

\begin{table}[htbp] 
        \centering
        \centering
        \renewcommand{\arraystretch}{1.2}
        \begin{tabular}{|>{\centering\arraybackslash}p{1.5cm}|>{\centering\arraybackslash}p{1.5cm}|>{\centering\arraybackslash}p{1.5cm}|>{\centering\arraybackslash}p{1.5cm}|}
          \hline
          group $g$ & index $\xi$ &  A & B    \\
          \hline \multirow{9}{*}{1} 
          & 1         &  (1,2,3)       & (4,5,6) \\\cline{2-4}
          & 2         &  (2,3,4)       & (1,5,6) \\\cline{2-4}
          & 3         &  (3,4,5)       & (1,2,6) \\\cline{2-4}
          & 4         &  (1,2,4)       & (3,5,6) \\\cline{2-4}
          & 5         &  (1,3,4)       & (2,5,6) \\\cline{2-4}
          & 6         &  (3,4,6)       & (1,2,5) \\\cline{2-4}
          & 7         &  (2,3,5)       & (1,4,6) \\\cline{2-4}
          & 8         &  (2,4,5)       & (1,3,6) \\\cline{2-4}
          & 9         &  (1,4,5)       & (2,3,6) \\
          \hline
        \end{tabular}
        \caption{
        Bi-partition patterns of the six-site cluster of the honeycomb lattice into two three-site subsystems A and B for calculating the EE. 
        The indices in A and B correspond to the site numbers in Fig.~\ref{Fig:honeycomb}(c).
}
        \label{table:honeycomb}
\end{table}

Figure~\ref{Fig:honeycomb}(a) shows the optimization process of the objective function \(L\). 
The changes in \(\lambda\) in Eq.~(\ref{eq:L}) and the learning rate \(\eta\) are shown in the lower panel. 
We note that $L$ is reduced faster by taking $\lambda$ nonzero on the way of the optimization than by taking it nonzero from the beginning.
The inset shows the changes of \(\bar{S}^T\) and \(\Delta S^T\); 
\(\bar{S}^T\) reaches around $1.7$, while \(\Delta S^T\) reaches a sufficiently small value. 
This suggests that the quantum entanglement is optimized with being almost identical for all bi-partition patterns. 
Note that \(\Delta S^T\) remains larger when optimizing with \(\lambda = 0\), leading to the spatially inhomogeneous interactions.

Figure~\ref{Fig:honeycomb}(b) displays the evolution of \(J_{ij}^\mu\) during the optimization process. 
One can observe that they eventually converge to one of the values $0$, $1$, or $-1$. 
Since each bond satisfies the condition \(\sum_\mu ({J_{ij}^{\mu}})^2 = 1\), this means that only one of the $\mu \in \{x,y,z\}$ becomes nonzero for each $J_{ij}$, resulting in an Ising-type anisotropic interaction. 
The optimized interactions of each bond are plotted by color on the lattice in Fig.~\ref{Fig:honeycomb}(c), with the color code in the inset. 
Note that we plot the absolute values of the optimized $J_{ij}$ since their signs do not matter to the value of the objective function; 
we confirm that changing all these signs to negative does not change the value of the objective function.
Interestingly, the interactions are all Ising-type and bond dependent, realizing the Kitaev model, which is known to exhibit spin liquid behavior with fractional Majorana excitations enabling topological quantum computations~\cite{KITAEV20062}.

\begin{figure*}[htbp]
        \centering
        \includegraphics[width=2.0\columnwidth,pagebox=cropbox,trim={60 50 100 40},clip]{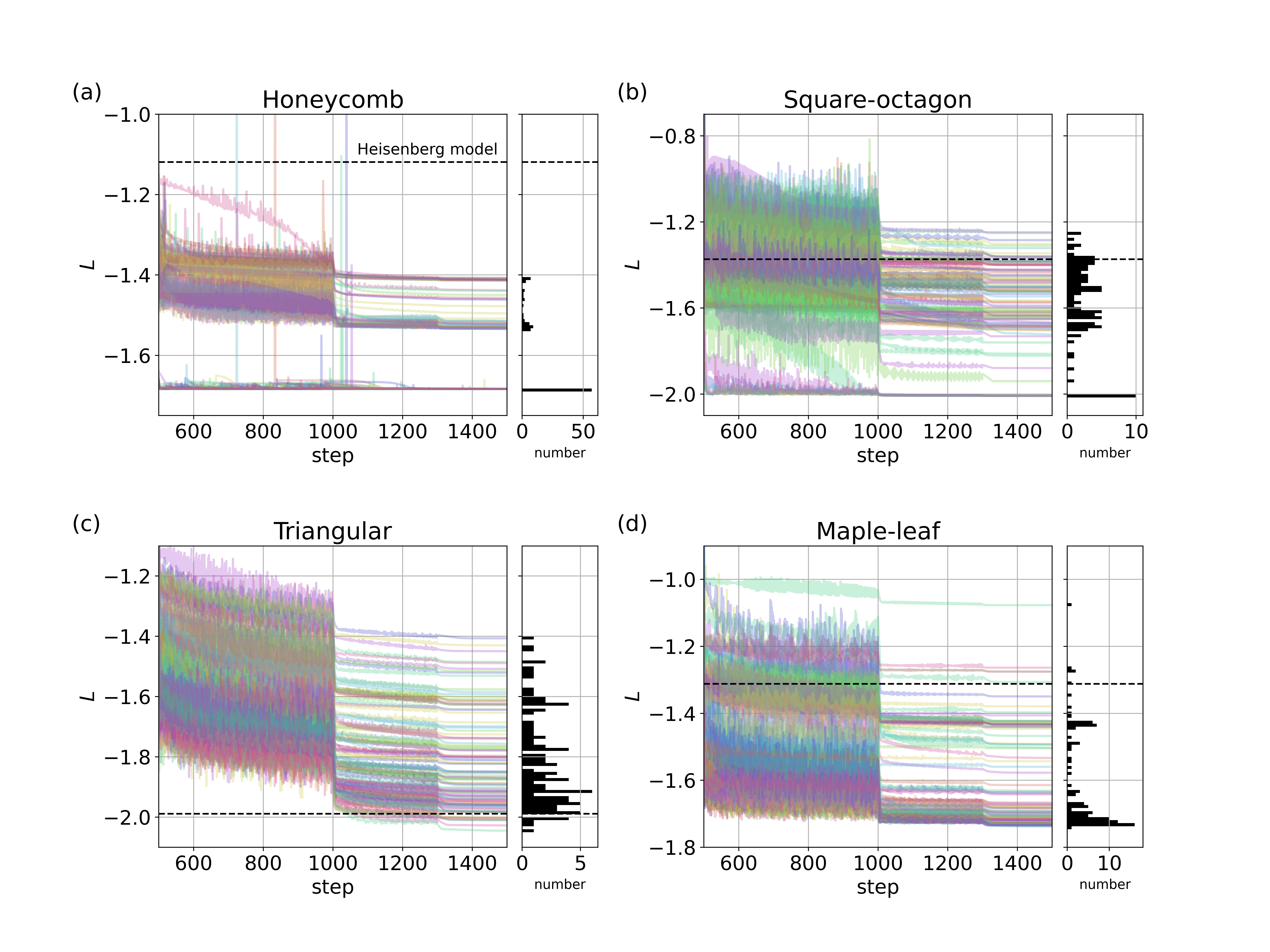}
        \caption{
        Changes in $L$ for 100 different initial conditions for (a) honeycomb, (b) square-octagon, (c) triangular, and (d) maple-leaf lattices.
        The right panel in each figure is the histogram of $L$ after convergence. 
        The dotted line indicates $L$ for the AFM Heisenberg model on each lattice.
        Note that the abrupt changes at steps 1000 and 1300 are due to the changes in the learning rate $\eta$, shown below in Fig.~\ref{Fig:honeycomb}(a). 
}
        \label{Fig:losses}
\end{figure*}

In Fig.~\ref{Fig:losses}(a), we present the optimization process of \(L\) for 100 different initial conditions. 
The histogram of the values of \(L\) at the end of the optimization is shown in the panel on the right. 
For comparison, we also show the value of $L$ for the antiferromagnetic (AFM) Heisenberg model with $J_{ij}^\mu = 1/\sqrt{3}$ for all $\langle ij \rangle$ and $\mu$, which is expected to show an AFM long-range order on this bipartite lattice. 
We find that all the 100 results have much smaller values of $L$ than this value. 
We also find that 57 out of 100 converge to the Kitaev model with the smallest value of $L$. 
In addition, there is a gap between the smallest value of $L$ and other values, indicating that the solutions of the Kitaev model are uniquely selected out.

Thus, the inverse design by maximizing the TEEE automatically finds the celebrated Kitaev model on the honeycomb lattice. 
The Kitaev model has the so-called bond frustration, where the local energy-optimal configuration cannot be satisfied on the whole due to the bond-dependent Ising-type anisotropic interactions~\cite{RevModPhys.87.1}.
This makes the wave function spatially nonlocal, leading to large quantum entanglement in the spin-liquid ground state.

\subsection{Square-octagon lattice}
\label{subsec:48}

\begin{figure}[htbp] 
        \centering
        \includegraphics[width=0.9\columnwidth,pagebox=cropbox,trim={100 130 100 120},clip]{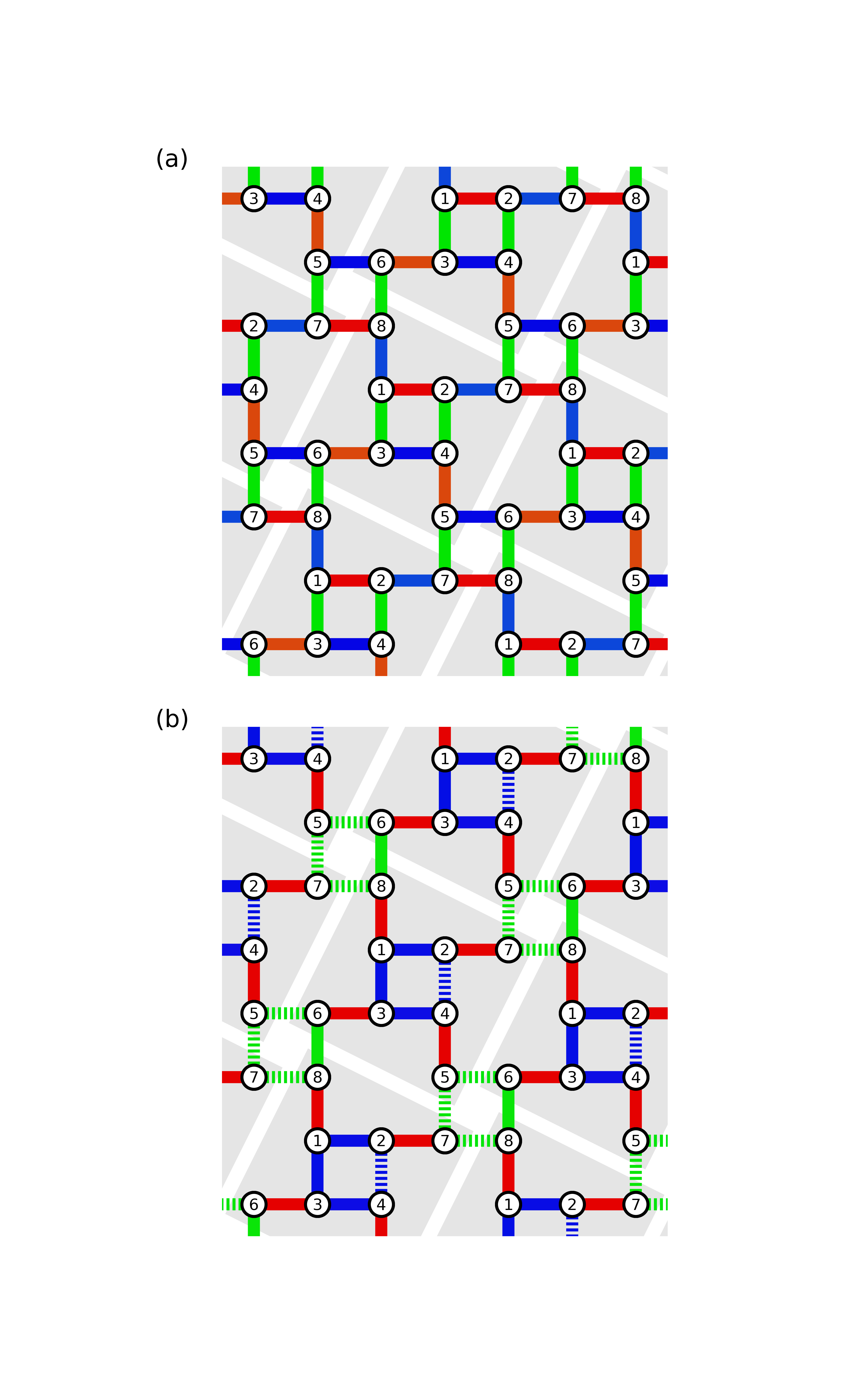} 
        \caption{
        Representative optimized interaction parameters for the square-octagon lattice model in the solution with (a) $L\simeq-2.0070$ and (b) $L\simeq-2.0086$, which are nondegenerate and doubly degenerate states, respectively. 
        The shades denote eight-site cluster unit cells. 
        The color of the bonds represents the value of each spin interaction according to the inset in Fig.~\ref{Fig:honeycomb}(c) and the solid (dotted) line indicates that the sign of the largest component of $J_{ij}^{\mu}$ is negative (positive).
        Note that though the optimized values of $J_{ij}^{\mu}$ have a mixture of both positive and negative values, we illustrate modified signs since the value of $L$ remains consistent in both cases: (a) when all these signs are either entirely positive or entirely negative and (b) when an odd number of bonds forming the square are positive.
}
        \label{Fig:48}
\end{figure}

Next, we apply the method to the square-octagon lattice, which is $1/5$-depleted square lattice, as illustrated in Fig.~\ref{Fig:48}. 
This lattice structure is realized in real materials~\cite{doi:10.1143/JPSJ.64.2758}, and has also been studied as a surface code in quantum computing~\cite{PhysRevLett.116.050501,PhysRevB.97.205404}.
We consider an eight-site cluster with periodic boundary conditions, comprising 12 distinct bonds and 36 parameters. 
We restrict our consideration to bi-partitions where the system is divided into two interconnected subsystems, with each subsystem having four sites, as detailed in Table \ref{table:48}. 
In contrast to the honeycomb case in Table \ref{table:honeycomb}, there are five groups, each with one to four elements. 
The inverse temperature is maintained at \(\beta=160\), which is larger than that for the honeycomb lattice case, because the larger cluster size increases the number of eigenstates.

\begin{table}[htbp] 
        \centering
        \centering
        \renewcommand{\arraystretch}{1.2}
        \begin{tabular}{|>{\centering\arraybackslash}p{1.5cm}|>{\centering\arraybackslash}p{1.5cm}|>{\centering\arraybackslash}p{1.5cm}|>{\centering\arraybackslash}p{1.5cm}|}
          \hline
          group $g$ & index $\xi$ &  A & B    \\
          \hline \multirow{1}{*}{1} 
          & 1  & (1,2,3,4) & (5,6,7,8) \\
          \hline \multirow{2}{*}{2} 
          & 1  & (1,2,7,8) & (3,4,5,6) \\\cline{2-4}
          & 2  & (1,3,6,8) & (2,4,5,7) \\
          \hline \multirow{4}{*}{3} 
          & 1  & (1,2,4,7) & (3,5,6,8) \\\cline{2-4}
          & 2  & (1,2,3,8) & (4,5,6,7) \\\cline{2-4}
          & 3  & (1,3,4,6) & (2,5,7,8) \\\cline{2-4}
          & 4  & (1,6,7,8) & (2,3,4,5) \\
          \hline \multirow{4}{*}{4} 
          & 1  & (1,2,3,7) & (4,5,6,8) \\\cline{2-4}
          & 2  & (1,2,4,5) & (3,6,7,8) \\\cline{2-4}
          & 3  & (1,5,7,8) & (2,3,4,6) \\\cline{2-4}
          & 4  & (1,3,4,8) & (2,5,6,7) \\
          \hline \multirow{4}{*}{5} 
          & 1  & (1,5,6,8) & (2,3,4,7) \\\cline{2-4}
          & 2  & (1,2,4,8) & (3,5,6,7) \\\cline{2-4}
          & 3  & (1,2,3,6) & (4,5,7,8) \\\cline{2-4}
          & 4  & (1,3,4,5) & (2,6,7,8) \\
          \hline
        \end{tabular}
        \caption{
        Bi-partition patterns for of eight-site cluster of the square-octagon lattice into two four-site subsystems A and B for calculating the EE. 
        The indices in A and B correspond to the site numbers in Fig.~\ref{Fig:48}.
}
        \label{table:48}
\end{table}

In Fig.~\ref{Fig:losses}(b), the optimization process of \(L\) is depicted for 100 different initial conditions. 
The value of $L$ of the AFM Heisenberg model is also plotted as a comparison, which has a lower value than the honeycomb lattice case and even lower than some optimized solutions. 
However, akin to the honeycomb lattice case, out of the 100 solutions, 10 converge to the minimal \(L\) value, distinctly separated from the other solutions by a discernible finite gap.

Figures~\ref{Fig:48}(a) and \ref{Fig:48}(b) illustrate two of these optimal solutions. 
The color scheme is the same as the inset of Fig.~\ref{Fig:honeycomb}(c). 
Remarkably, all the bonds shown in both cases exhibit bond-dependent Ising-type anisotropy. 
The solution in Fig.~\ref{Fig:48}(a) is nondegenerate, while that in Fig.~\ref{Fig:48}(b) is doubly degenerate. 
Note that we modify the signs of $J_{ij}^\mu$ in Figs.~\ref{Fig:48}, following the rules that we found:
In Fig.~\ref{Fig:48}(a), the value of \(L\) remains consistent whether all signs of bonds are uniformly negative or uniformly positive. 
Conversely, in Fig.~\ref{Fig:48}(b), \(L\) remains the same value if an odd number of the four bonds that form a square are positive, with the remaining bonds being negative; the signs of any other bonds do not affect the value of \(L\).


We note that some bonds are not completely anisotropic, e.g., \(J_{36} \sim 0.96\) in Fig.~\ref{Fig:48}(a). 
However, using the purification method outlined in Appendix~\ref{sec:S_params}, they are transformed into fully anisotropic Hamiltonians with keeping the value of \(L\) almost intact. 
The remaining eight optimal solutions are almost identical to either of the two solutions.

The result in Fig.~\ref{Fig:48}(a) is known as the Kitaev model, which manifests a spin liquid in the ground state~\cite{PhysRevB.76.180404,Kells_2011}. 
It is understood that this has large entanglement due to the bond frustration, like the honeycomb lattice case. 
Note that the solution in Fig.~\ref{Fig:48}(b) is different from the Kitaev model, but it also has bond frustration due to the anisotropic interactions with different signs in the four-bond squares as mentioned above.

\subsection{Triangular lattice}
\label{subsec:tri}

\begin{figure}[htbp]
        \centering
        \includegraphics[width=1.0\columnwidth,pagebox=cropbox,trim={120 60 120 60},clip]{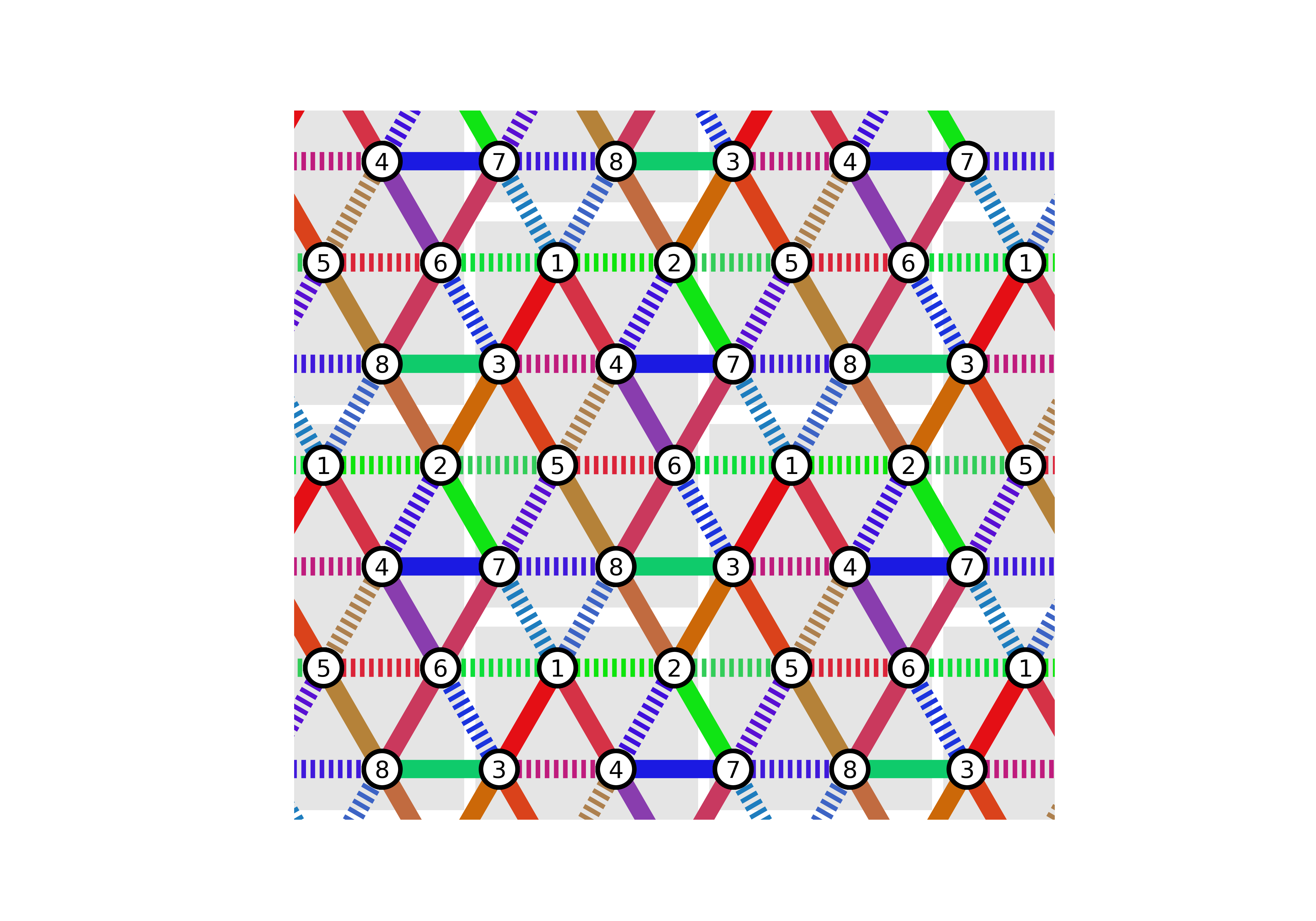}
        \caption{
        Representative optimized interaction parameters for the triangular lattice model in the solution with the lowest $L\simeq-2.0451$. 
        The shades denote eight-site cluster unit cells. 
        The notations are common to those in Fig.~\ref{Fig:48}. 
}
        \label{Fig:triangular}
\end{figure}

Here, we apply the method to the triangular lattice known to have geometrical frustration.
The quantum spin models with geometrical frustration have been studied as a possible route to spin liquid states, such as the resonating valence bond state~\cite{Balents2010}. 
We consider eight-site cluster with periodic boundary conditions as illustrated in Fig.~\ref{Fig:triangular}. 
This comprises 24 distinct bonds and 72 parameters. 
We restrict our consideration to bi-partitions where the system is divided into two interconnected subsystems, with each subsystem having four sites, as detailed in Table \ref{table:tri}. 
The inverse temperature is maintained at \(\beta=160\).

\begin{table}[htbp] 
        \centering
        \centering
        \renewcommand{\arraystretch}{1.2}
        \begin{tabular}{|>{\centering\arraybackslash}p{1.5cm}|>{\centering\arraybackslash}p{1.5cm}|>{\centering\arraybackslash}p{1.5cm}|>{\centering\arraybackslash}p{1.5cm}|}
          \hline
          group $g$ & index $\xi$ &  A & B    \\
          \hline \multirow{8}{*}{1} 
          & 1 & (1,2,3,4) &  (5,6,7,8) \\\cline{2-4}
          & 2 & (1,2,3,8) &  (4,5,6,7) \\\cline{2-4}
          & 3 & (1,6,7,8) &  (2,3,4,5) \\\cline{2-4}
          & 4 & (1,3,6,8) &  (2,4,5,7) \\\cline{2-4}
          & 5 & (1,2,4,7) &  (3,5,6,8) \\\cline{2-4}
          & 6 & (1,3,4,6) &  (2,5,7,8) \\\cline{2-4}
          & 7 & (1,4,6,7) &  (2,3,5,8) \\\cline{2-4}
          & 8 & (1,2,7,8) &  (3,4,5,6) \\
          \hline \multirow{4}{*}{2} 
          & 1 & (1,3,4,5) & (2,6,7,8) \\\cline{2-4}
          & 2 & (1,2,4,8) & (3,5,6,7) \\\cline{2-4}
          & 3 & (1,3,6,7) & (2,4,5,8) \\\cline{2-4}
          & 4 & (1,5,7,8) & (2,3,4,6) \\
          \hline \multirow{4}{*}{3} 
          & 1 & (1,2,3,7) & (4,5,6,8) \\\cline{2-4}
          & 2 & (1,4,6,8) & (2,3,5,7) \\\cline{2-4}
          & 3 & (1,3,5,8) & (2,4,6,7) \\\cline{2-4}
          & 4 & (1,4,5,7) & (2,3,6,8) \\
          \hline
        \end{tabular}
        \caption{
        Bi-partition patterns of eight-site cluster of the triangular lattice into two four-site subsystems A and B for calculating the EE. 
        The indices in A and B correspond to the site numbers in Fig.~\ref{Fig:triangular}. 
}
        \label{table:tri}
\end{table}

Figure~\ref{Fig:losses}(c) shows the optimization processes for 100 different initial conditions. 
There is only one solution that gives the minimum value of $L$, and several different solutions appear without showing an apparent gap, forming rather continuous spectrum. 
This is in stark contrast to the cases of honeycomb and square-octagon lattices where multiple results starting from different initial conditions converge to the specific optimal solutions well separated from the others. 
In addition, the value of $L$ for the AFM Heisenberg model is close to the lowest $L$ solution, which is also in contrast to the previous two cases. 
The ground state of the AFM Heisenberg model on the triangular lattice is expected to exhibit a noncollinear $120^{\circ}$ order~\cite{PhysRevB.50.10048,PhysRevLett.82.3899}, where the magnetic moment $\sim 0.20$~\cite{PhysRevLett.82.3899,PhysRevB.74.224420} is reduced compared to that for the honeycomb lattice case $\sim 0.27$~\cite{PhysRevB.73.054422}. 
This suggests larger quantum entanglement in the triangular lattice case than the honeycomb lattice case. 
Notably, our method can generate models with larger entanglement than the AFM Heisenberg model. 
In Fig.~\ref{Fig:triangular}, we plot the optimized interactions for the solution with the lowest $L$ value in Fig.~\ref{Fig:losses}(c). 
The interactions are spatially inhomogeneous without showing any clear pattern, but the anisotropy is relatively prevalent. 


The lack of convergence to a single solution starting from different initial conditions in the triangular lattice case can be attributed to possible coexistence of two different types of frustration, namely, geometrical frustration and bond frustration. 
In contrast to the bipartite honeycomb and square-octagon with no geometrical frustration, the triangular lattice is non-bipartite and can exploit both two frustrations, for which multiple metastable solutions may arise rather than converge to a single solution.
Indeed, the representative solution in Fig.~\ref{Fig:triangular} appears to reconcile two frustrations by adopting the spatially-inhomogeneous anisotropic interactions.

\subsection{Maple-leaf lattice}
\label{subsec:maple}

\begin{figure}[htbp]
        \centering
        \includegraphics[width=1.0\columnwidth,pagebox=cropbox,trim={110 120 110 100},clip]{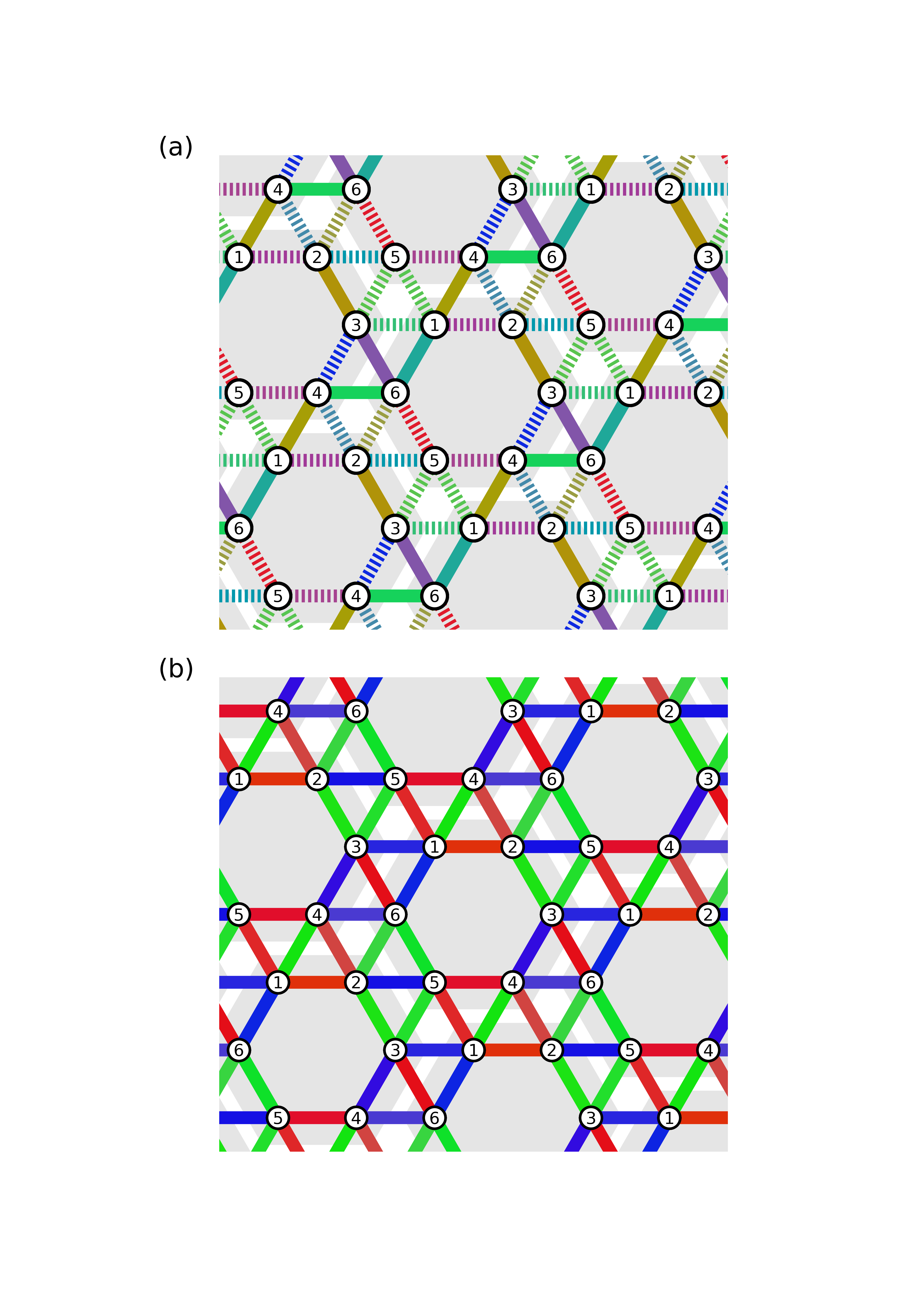}
        \caption{
        Representative optimized interaction parameters for the maple-leaf lattice model in the solutions with (a) $L\simeq-1.738$ and (b) $L\simeq-1.708$. 
        The shades denote six-site cluster unit cells. 
        The notations are common to those in Fig.~\ref{Fig:48}. 
        In (a), the signs appear to be random, while in (b) the signs are all taken to be negative though the optimized solutions have a mixture of both positive and negative, since the value of $L$ remains consistent in both cases. 
}
        \label{Fig:mapleleaf}
\end{figure}

Finally, we apply the method to the maple-leaf lattice, which is a $1/7$-depleted triangular lattice, and hence, has a slightly weaker geometrical frustration than the triangular lattice.
Quantum spin systems with this lattice structure are realized in real materials~\cite{Fennell_2011,https://doi.org/10.1002/anie.201203775}. 
We consider a six-site cluster with periodic boundary conditions, as illustrated in Fig.~\ref{Fig:mapleleaf}. 
This lattice comprises 15 distinct bonds and 45 parameters. 
We restrict our consideration to bi-partitions where the system is divided into two interconnected subsystems, with each subsystem having three sites, as detailed in Table \ref{table:maple}. 
The inverse temperature is maintained at \(\beta=60\).

\begin{table}[htbp] 
        \centering
        \centering
        \renewcommand{\arraystretch}{1.2}
        \begin{tabular}{|>{\centering\arraybackslash}p{1.5cm}|>{\centering\arraybackslash}p{1.5cm}|>{\centering\arraybackslash}p{1.5cm}|>{\centering\arraybackslash}p{1.5cm}|}
          \hline
          group $g$ & index $\xi$ &  A & B    \\
          \hline \multirow{3}{*}{1} 
          & 1 & (1,2,3) &  (4,5,6) \\\cline{2-4}
          & 2 & (1,2,6) &  (3,4,5) \\\cline{2-4}
          & 3 & (1,5,6) &  (2,3,4) \\
          \hline \multirow{6}{*}{2} 
          & 1 & (1,2,5) &  (3,4,6) \\\cline{2-4}
          & 2 & (1,4,5) &  (2,3,6) \\\cline{2-4}
          & 3 & (1,3,4) &  (2,5,6) \\\cline{2-4}
          & 4 & (1,3,6) &  (2,4,5) \\\cline{2-4}
          & 5 & (1,2,4) &  (3,5,6) \\\cline{2-4}
          & 6 & (1,4,6) &  (2,3,5) \\
          \hline \multirow{1}{*}{3} 
          & 1 & (1,3,5) &  (2,4,6) \\
          \hline
        \end{tabular}
        \caption{
        Bi-partition patterns of the six-site cluster of the maple-leaf lattice into two three-site subsystems A and B for calculating the EE. 
        The indices in A and B correspond to the site numbers in Fig.~\ref{Fig:mapleleaf}. 
}
        \label{table:maple}
\end{table}

As depicted in Fig.~\ref{Fig:losses}(d), there is only one solution with the smallest $L$, without showing a clear gap to the other solutions, among the calculations for 100 initial conditions. 
This is similar to the triangular lattice case, but also different from the results of the honeycomb and square-octagon lattices, where multiple calculations converge to a single optimal solution.
We plot the result for the smallest value of $L\simeq-1.738$ in our calculations in Fig.~\ref{Fig:mapleleaf}(a). 
It appears that the interactions are asymmetric and spatially inhomogeneous.
Furthermore, the signs of the interactions do not show a specific spatial pattern. 
This may be due to the fact that, similarly to the triangular lattice case, both geometrical frustration and bond frustration are allowed to coexist. 
Meanwhile, $L$ of the AFM Heisenberg model is not that small, as shown in Fig.~\ref{Fig:losses}(d), presumably because the maple-leaf lattice has weaker geometrical frustration than the triangular lattice.

We also plot the solution with a larger $L\simeq-1.708$ in Fig.~\ref{Fig:mapleleaf}(b). 
Contrary to the other solutions, it has a unique structure constructed solely by Ising-type anisotropic interactions.
The structure includes the quadrangle by four $J^x$ ($J^y$, $J^z$) enclosing $J^y$ ($J^z$, $J^x$), and is symmetric for not only $C_2$ rotation but also operations that simultaneously perform a $C_3$ rotation in real space and a $2\pi/3$ rotation about the [111] axis in spin space.
Although the results in Fig.~\ref{Fig:mapleleaf}(b) have a mixture of both positive and negative values for the value of \(J_{ij}^{\mu}\), we confirm that changing all these signs to negative does not change the value of $L$. 
Thus, although this solution is not the best one in terms of quantum entanglement, it gives a concise model with bond-dependent Ising-type interactions compatible with the structural unit cell. 
To the best of our knowledge, such a model has not been reported thus far.  
This demonstrates that our framework is useful to develop an unprecedented, interesting model.

\section{Discussion}
\label{sec:discussion}

In Sec.~\ref{sec:result}, we showed that our method successfully finds a model with optimized TEEE for each lattice. 
Interestingly, the solutions look qualitatively different between the bipartite lattices without geometrical frustration (honeycomb and square-octagon) and the non-bipartite lattices with geometrical frustration (triangular and maple-leaf). 
For the former, the optimization processes starting from different initial conditions converge onto a single solution with strongly anisotropic Kitaev-type interactions, while for the latter, they do not appear to single out a unique solution. 
Nevertheless, the interesting observation is that, even for the latter geometrically frustrated cases, the obtained solutions are not isotropic but rather anisotropic in spin space; the optimized TEEE has a lower value than that for the isotropic AFM Heisenberg model. 
This suggests that the systems try to maximize the quantum entanglement by exploiting the bond frustration even in the presence of geometrical frustration. 

\begin{figure}[htbp]
        \centering
        \includegraphics[width=1.0\columnwidth,pagebox=cropbox,clip]{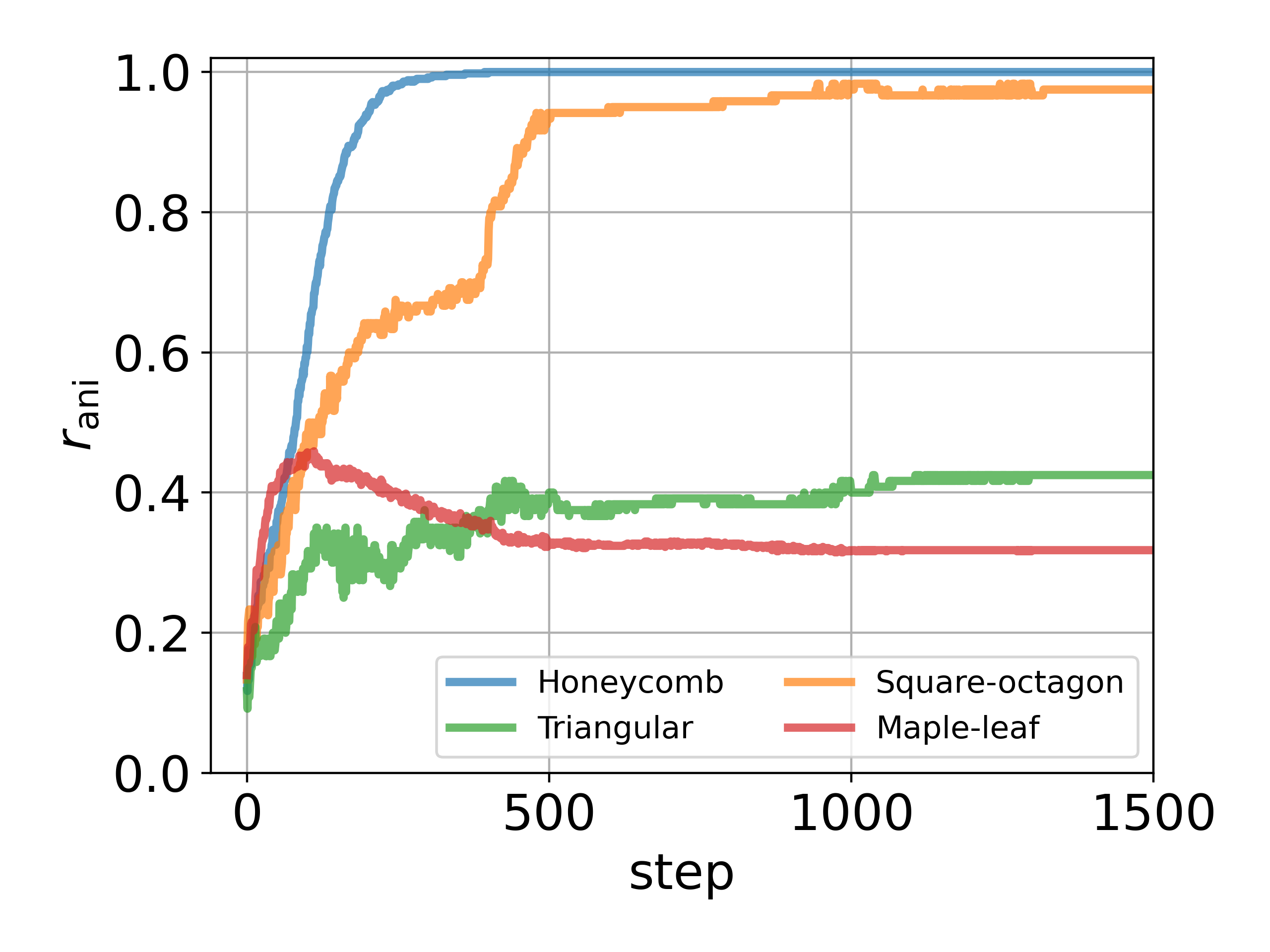}
        \caption{
        Changes in the ratio of anisotropic spin interactions, $r_{\rm ani}$ in Eq.~(\ref{eq:aniso}), during the optimization process on each lattice.
}
        \label{Fig:J_directed}
\end{figure}

For more comparison, we estimate the degree of spin anisotropy in each lattice by computing the ratio of the anisotropic interactions by
\begin{align}
r_{\rm ani} = \left\langle \frac{1}{N_b} \sum_{\langle ij \rangle} \sum_\mu f(J_{ij}^{\mu}) \right\rangle,  \label{eq:aniso}
\end{align}
where $f(x)=1$ for $|x|>x_{\rm thres}$ and otherwise $f(x)=0$ (we take the threshold $x_{\rm thres}=0.95$ in the following calculations);
$N_b$ is the number of distinct bonds in each cluster used in the calculations, and the average $\langle \cdots \rangle$ is taken over solutions with $L$ in the range satisfying $L < 0.98L_{\rm min}$ among the 100 calculations, where $L_{\rm min}$ is the lowest value of $L$. 
The results are plotted in Fig.~\ref{Fig:J_directed}. 
For the honeycomb and square-octagon lattices that are free from geometrical frustration, the values of $r_{\rm ani}$ are rapidly increased;
$r_{\rm ani}$ reaches almost $1$ in the honeycomb lattice case, and it exceeds $0.95$ in the square-octagon case. 
In contrast, the increase of $r_{\rm ani}$ is less pronounced in the triangular and maple-leaf lattices with geometrical frustration; 
$r_{\rm ani}$ is about 0.4 in the triangular lattice case, and it once increases but decreases down to $\sim 0.3$ in the maple-leaf lattice. 
Nonetheless, the final values of $r_{\rm ani}$ are nonzero and larger than those for the initial states with random $J_{ij}^\mu$, indicating that the optimization of TEEE favors anisotropic interactions even for the geometrically frustrated cases. 
This is an important lesson from the automatic design: To optimize the quantum entanglement, it would be, in general, helpful to utilize anisotropic interactions. 
This is rather counterintuitive, since the anisotropic interactions suppress spin-singlet formation that is widely recognized to be crucial to stabilize quantum spin liquids like the resonating valence bond state.

\section{Conclusion and outlook}
\label{sec:conclusion}

In summary, we have developed a method to design the Hamiltonian with large quantum entanglement using automatic differentiation. 
In the method, the parameters in Hamiltonians are optimized so as to reduce the objective function corresponding to the desired entanglement property of the system. 
Among several quantities representing the quantum entanglement, we introduced an extension of the entanglement entropy, which we call the thermal ensemble of entanglement entropy, as the objective function, to achieve stable and efficient optimization. 
Applying this method to the quantum spin systems on four different lattice structures, we showed that it successfully generates the Hamiltonian with large quantum entanglement. 
The optimization on the honeycomb and square-octagon lattices automatically finds the Kitaev model, which is known to provide exactly-solvable spin liquid ground states. 
On the triangular and maple-leaf lattices with geometrical frustration, the results do not converge to a specific model, suggesting a degenerate manifold of the models with spatially inhomogeneous interactions. 
Through the optimization, however, we found an unprecedented homogenous model with large entanglement for the maple-leaf lattice case. 
Furthermore, by comparing optimized interactions on these different lattices, we found that the introduction of anisotropic interactions plays an important role in the enhancement of quantum entanglement even in the presence of geometrical frustration.
This finding is rather incompatible with the conventional wisdom that the spin-singlet formation by isotropic Heisenberg interaction is widely recognized as a key ingredient for the realization of quantum entangled states, such as the resonating valence bond state.

The optimized solutions in Sec.~\ref{subsec:honeycomb} correspond to the gapless phase of the Kitaev model. 
This model is exactly-solvable and the ground state offers a quantum spin liquid with two types of fractional excitations, itinerant Majorana fermions and localized $Z_2$ fluxes~\cite{KITAEV20062}. 
The entanglement entropy of this model was shown to have two separate contributions from the two fractional excitations~\cite{PhysRevLett.105.080501}. 
It is noteworthy that the framework automatically designs such an intriguing model with exact solvability. 
Meanwhile, the inverse design discovered new models on the square-octagon and maple-leaf lattices. 
These models may also have interesting properties of quantum entanglement. 
This is a subject for future research.


Our approach has a high flexibility, being effectively applicable to a diverse range of systems, including quantum spin systems, strongly correlated systems for both fermionic and bosonic cases, and even non-Hermitian systems. 
Nonetheless, an increase in the number of parameters tends to complicate the process of attaining optimal solutions. 
To address this issue, it might be necessary to develop new strategies for the efficient sampling of initial conditions, potentially employing techniques like Bayesian optimization. 
However, it is important to note that finding a globally optimal solution is not a prerequisite for the successful identification of unprecedented models, as exemplified by the finding of new models on the square-octagon and maple-leaf lattice cases. 
Therefore, despite the potential challenges in resolving global optimal issues, the current approach retains substantial utility for the discovery of new models in various complex systems.

In this research, the exact diagonalization was employed to calculate the entanglement entropy. 
However, the automatic differentiation of exact diagonalization is notably memory-intensive. 
When utilizing a single NVIDIA A100 GPU machine with 40 GB of memory, the maximum manageable matrix size is \(2^{12} \times 2^{12}\). 
Given the versatility of automatic differentiation, it can be integrated with the Lanczos method or tensor networks to accommodate larger system sizes. 
It is noteworthy that the implementation of automatic differentiation in tensor networks has been documented recently~\cite{PhysRevX.9.031041}.

Exploring applications to diverse entanglement measures promises to be intriguing. 
Maximizing of topological entanglement entropy could pave the way for novel topological quantum systems. 
For instance, applying this to the out-of-time-ordered correlation~\cite{larkin1969quasiclassical,Shenker2014,Maldacena2016} and tripartite mutual information~\cite{PhysRevA.97.042330} might unveil distinctive quantum scrambling behaviors. 
Historically, the conception of models and systems has predominantly relied on the experience and intuition of researchers. 
The automation of system construction, possessing the target quantum functionality through an inverse design approach, would facilitate the uncovering of innovative principles and systems.

\begin{acknowledgements}
The authors thank Motoki Amano, Keisuke Fujii, Yasuyuki Kato, Kaito Kobayashi, Rico Pohle, Eiji Saitoh, Kotaro Shimizu,  Nobuyuki Yoshioka, and Junki Yoshitake for their valuable comments.  
They also thank Toshio Mori for his support of the computational environment. 
This work was supported by the Center of Innovations for Sustainable Quantum AI (JST Grant Number JPMJPF2221), 
MEXT Quantum Leap Flagship Program (MEXT Q-LEAP) Grant No. JPMXS0120319794, KAKENIHI Grant No. 20H00122, a Grant-in-Aid for Scientific Research on Innovative Areas “Quantum Liquid Crystals” (KAKENHI Grant No. JP19H05825) from JSPS of Japan, and JST CREST Grant No. JP-MJCR18T2.
\end{acknowledgements}

\appendix

\section{Mutual information and logarithmic negativity}
\label{sec:mutinfo}

\begin{figure}[htbp]
        \centering
        \includegraphics[width=1.0\columnwidth,pagebox=cropbox,trim={30 10 60 50},clip]{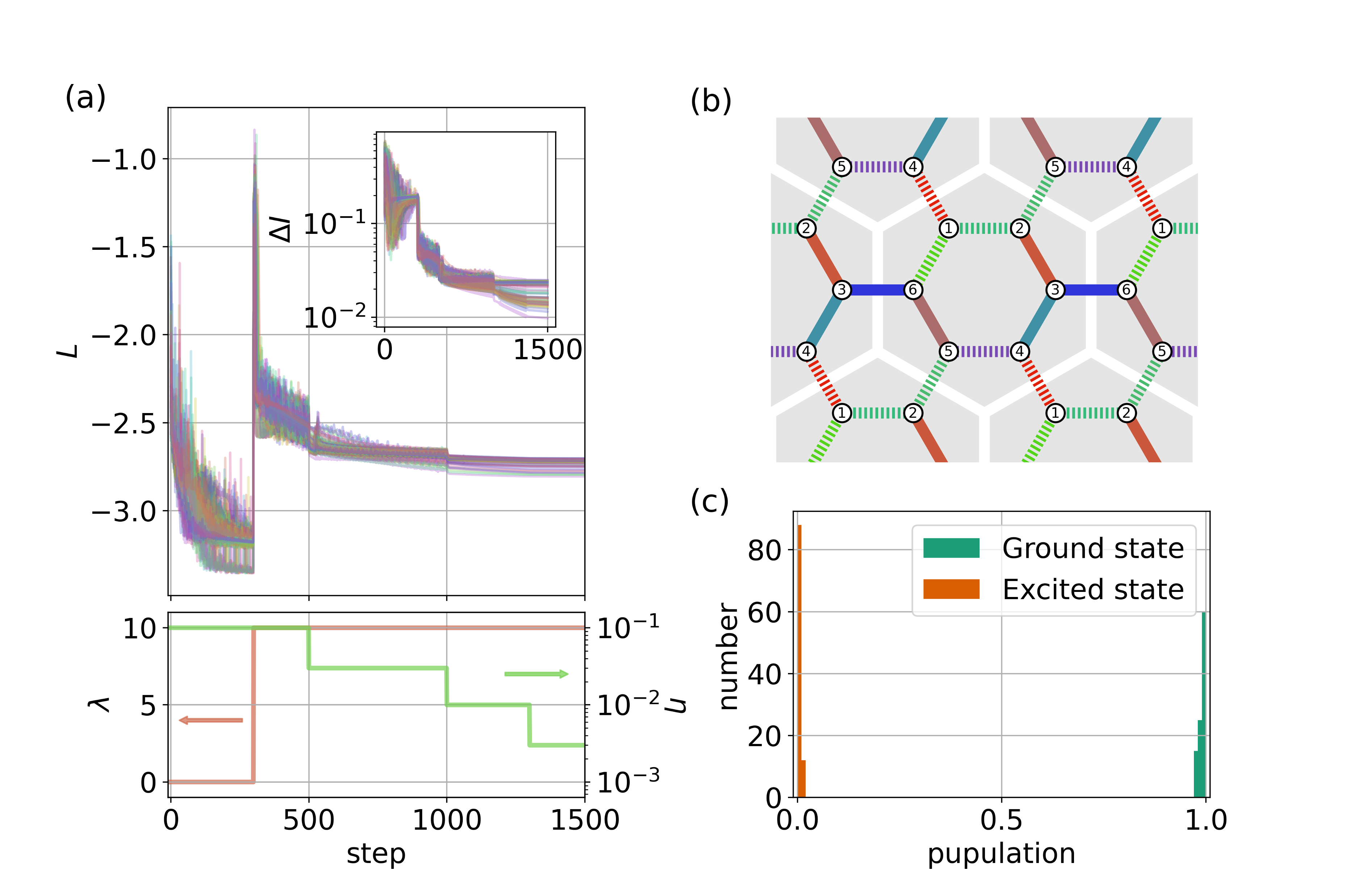}
        \caption{
        (a) Changes in $L$ using the MI for 100 different initial conditions for the model on a honeycomb lattice. 
        Note that the abrupt increase in $L$ at the 300 step is due to the change in $\lambda$. 
        The inset shows the changes in $\Delta I$. 
        The lower panel shows the schedule of $\lambda$ and learning rate $\eta$. 
        (b) Optimized interaction parameters. 
        The notations are common to those in Fig.~\ref{Fig:48}. 
        (c) Histogram of the populations of the ground state and the first excited state for 100 solutions. 
}
        \label{Fig:appendix_I}
\end{figure}

In this Appendix, we discuss the results of maximizing two different measures of quantum entanglement for mixed states, MI and LN. 
First, we discuss the MI, which is defined by 
\begin{align}
I &= S^{\rm B}_{\rm A} + S^{\rm B}_{\rm B} - S^{\rm B}_{\rm tot}, 
\label{eq:I}
\end{align}
where 
\begin{align}
S^{\rm B}_{\rm A(B)} &= -{\rm Tr} \rho_{\rm A(B)}^{\rm B} \log\rho_{\rm A(B)}^{\rm B}, \\
S^{\rm B}_{\rm tot} &= -{\rm Tr} \rho^{\rm B} \log\rho^{\rm B}, 
\end{align}
with 
\begin{align}
\rho_{\rm A(B)}^{\rm B} = -{\rm Tr_{\rm B(A)}} \rho^{\rm B} \log\rho^{\rm B}. 
\end{align}
Here, $\rho^{\rm B}$ is the thermal density matrix defined as 
\begin{align}
\rho^{\rm B}(\beta) = \frac{\sum_n e^{-\beta E_n} |\psi_n \rangle \langle \psi_n|}{\sum_n e^{-\beta E_n}}. 
\label{eq:neg}
\end{align}
The Hamiltonian and its bi-partitions are the same as in Sec.~\ref{subsec:honeycomb}.
The objective function $L$ is given by Eq.~(\ref{eq:L}) with the replacement of $S^T_{g,\xi}$ with $I_{g,\xi}$ in Eqs.~(\ref{eq:L})-(\ref{eq:barSg}), where $I_{g,\xi}$ is the MI for bi-partition with group $g$ and index $\xi$.
$\bar{I}$ and $\Delta I$ are defined accordingly. 
We set $\beta = 60$.

Figures~\ref{Fig:appendix_I}(a) shows the optimization process of $L$ for the calculations with 100 different initial conditions for the model in Eq.~(\ref{eq:Ham}) on the honeycomb lattice.
While the optimized values of $L$ exhibit similar magnitudes, they do not converge to a specific optimal state, which stands in stark contrast to the results for the TEEE in Fig.~\ref{Fig:losses}(a), where more than half of the solutions converge to the same optimal state. 
Furthermore, we note that the magnitudes of $\Delta I$ in this scenario, being on the order of $10^{-2}$ as shown in the inset of Fig.~\ref{Fig:appendix_I}(a), are considerably larger than those observed in the case of the TEEE, indicating that the optimized results are not spatially homogenous. 
Figure~\ref{Fig:appendix_I}(b) depicts the optimized interaction parameters for the solution with the smallest value of $L$. 
The result breaks both translational and rotational symmetry. 
We confirmed that most of the other solutions also have low symmetry compared with the result obtained in the optimization using the TEEE in Sec.~\ref{subsec:honeycomb}. 

Figure~\ref{Fig:appendix_I}(c) displays a histogram of the populations of the ground state and the first excited state in the 100 optimized results. 
The population of a state $|\psi \rangle$ is defined by $\langle \psi | \rho^{\rm B} | \psi \rangle$. 
The ground state clearly dominates the population distribution. 
This is attributed to the term $S^{\rm B}_{\rm tot}$ in Eq.~(\ref{eq:I}), which tries to make the thermal density matrix as close to a pure state as possible and avoid degeneracy in the ground state during the optimization process. 
Such a trend is in stark contrast to the optimization process for the TEEE that allows degeneracy. 
This is the reason why the TEEE has an advantage over the MI in the current problem; 
the interaction parameters with high spatial symmetry are unlikely to emerge when attempting to increase the MI because it avoids degeneracy.


\begin{figure}[htbp]
        \centering
        \includegraphics[width=1.0\columnwidth,pagebox=cropbox,trim={30 10 60 50},clip]{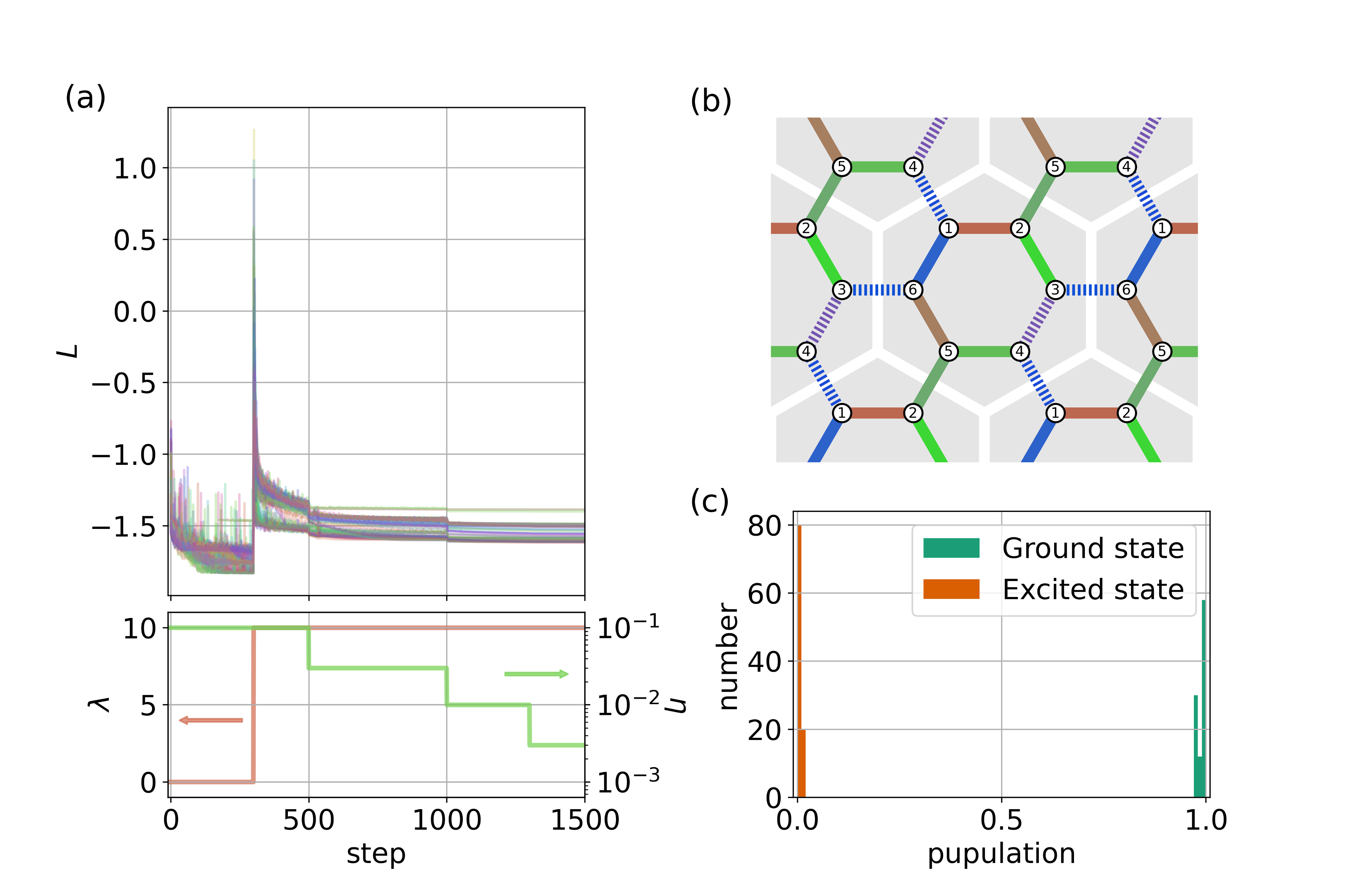}
        \caption{
        (a) Changes in $L$ using the LN for 100 different initial conditions for the model on a honeycomb lattice. 
        Note that the abrupt increase in $L$ at the 300 step is due to the change in $\lambda$. 
        The lower panel shows the schedule of $\lambda$ and learning rate $\eta$. 
        (b) Optimized interaction parameters. 
        The notations are common to those in Fig.~\ref{Fig:48}. 
        (c) Histogram of the population of the ground state and the first excited state for 100 solutions. 
}
        \label{Fig:appendix_logneg}
\end{figure}

Next, we discuss the results for another measure of quantum entanglement for mixed states, the LN, which is defined by 
\begin{align}
E^{\mathcal{N}} = \log || \rho^{\rm B}_{T_{\rm B}}(\beta) ||, 
\label{eq:neg}
\end{align}
where $||\cdot||$ denotes the trace norm of the matrix, and $\rho^{\rm B}_{T_{\rm B}}$ is the partial transpose over the subsystem B of $\rho^{\rm B}$. 
In the numerical calculation, $|| \rho^{\rm B}_{T_{\rm B}}(\beta) ||$ is computed by summing up the absolute values of the eigenvalues obtained by diagonalizing $\rho^{\rm B}_{T_{\rm B}}$.
Since the LN is an entanglement monotone~\cite{doi:10.1080/09500340008244048}, maximizing it can create a system with large entanglement.
The optimization is performed similarly to the MI case above. 

Figure~\ref{Fig:appendix_logneg} summarizes the results by using the LN in a similar manner to Fig.~\ref{Fig:appendix_I}. 
The results are qualitatively similar to those obtained by the MI; 
the optimization does not single out a specific optimal state from the others, the interaction parameters break the lattice symmetry, and the population is dominated by the ground state. 
Thus, we conclude that the LN is inferior to the TEEE as well as the MI.

\section{Parametrization dependence}
\label{sec:app_con}

\begin{figure}[htbp]
        \centering
        \includegraphics[width=1.0\columnwidth,pagebox=cropbox,trim={0 0 0 0},clip]{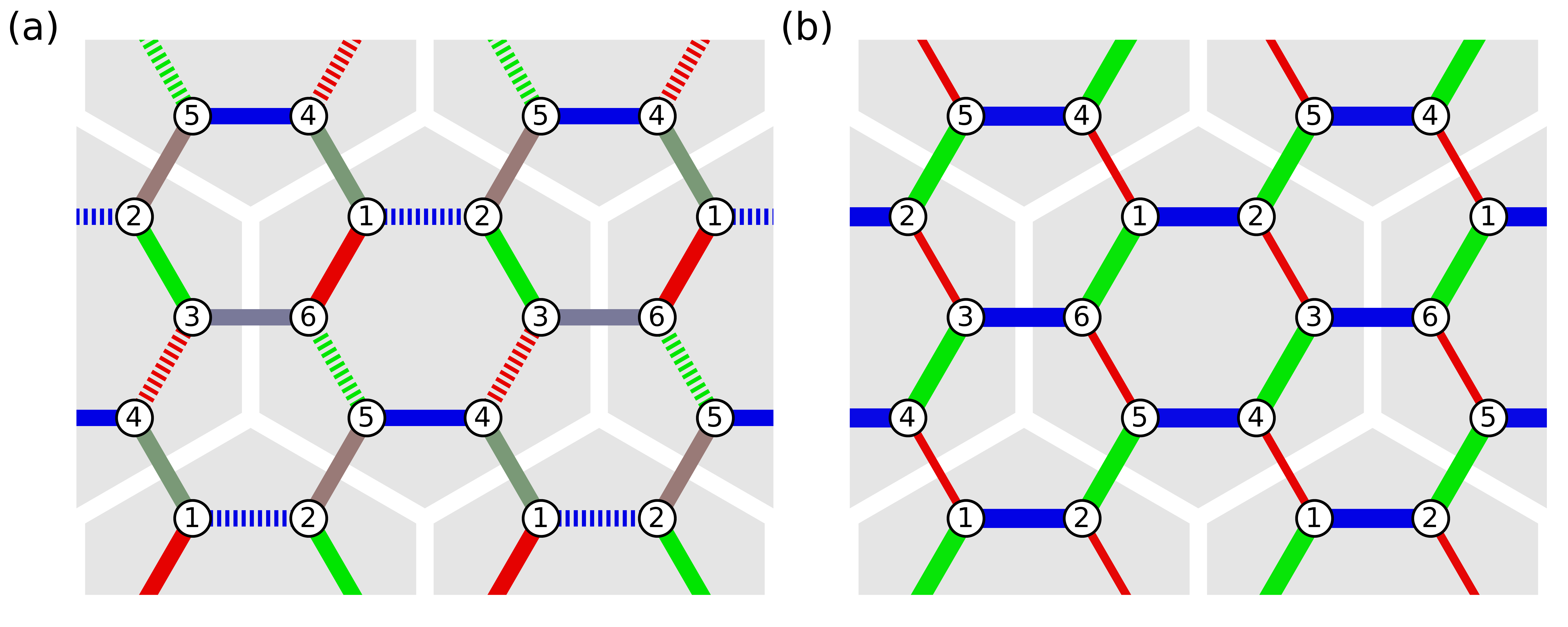}
        \caption{
        (a),(b) The optimized interaction parameters for the honeycomb lattice model with the use of the objective function of Eq.~(\ref{eq:L_LN}) under the constraint of (a) Eq.~(\ref{eq:Jij}) and (b) Eq.~(\ref{eq:Jij_2}). 
        The notations are common to those in Fig.~\ref{Fig:48}. 
        The thickness of the bond corresponds to the absolute value of its interaction. 
        See the text for details.
}
        \label{Fig:appendix_logneg_con}
\end{figure}

In this Appendix, we discuss how the way of parametrization affects the optimization. 
As mentioned in Sec.~\ref{subsec:calculationflow}, the parametrization in Eq.~\eqref{eq:Jij} is not unique. 
Instead, one can take, for instance, 
\begin{align}
J_{ij}^{\mu} = \frac{\sqrt{N_b}\theta_{ij}^\mu}{\sqrt{\sum_{\langle i,j\rangle \mu} {\theta_{ij}^\mu}^2}},  \label{eq:Jij_2}
\end{align}
where $N_b$ is the number of distinct bonds in each cluster used for the calculations. 
Note that Eq.~(\ref{eq:Jij}) satisfies the local constraint on each bond, $\sum_\mu (J_{ij}^\mu)^2=1$, while Eq.~(\ref{eq:Jij_2}) satisfies the global constraint, $\sum_{\langle i,j \rangle\mu} (J_{ij}^\mu)^2 = N_b$. 
Hence, Eq.~(\ref{eq:Jij_2}) allows different bond interactions to take different magnitudes; 
namely, it allows the optimization in wider parameter range than in Eq.~(\ref{eq:Jij}). 

The optimized results depend on not only the parametrization but also the objective function $L$. 
Indeed, we find that the results are almost intact for the parameterizations in Eq.~(\ref{eq:Jij}) and Eq.~(\ref{eq:Jij_2}) when taking the TEEE in Eq.~(\ref{eq:L}) for $L$, but depend on the parameterization when taking the LN in Eq.~(\ref{eq:neg}) for $L$. 
We demonstrate this below. 
Firstly, we perform calculations based on the parametrization in Eq.~(\ref{eq:Jij}), by using the objective function $L(\bm{\theta})$ defined as
\begin{align}
L(\bm{\theta}) = -\bar{E}^{\mathcal{N}}(\bm{\theta})  = -\frac{1}{\sum_g N_{g}} \sum_{g\xi} E^{\mathcal{N}}_{g,\xi}(\bm{\theta}) \label{eq:L_LN}, 
\end{align}
where $E^{\mathcal{N}}_{g,\xi}$ is the LN for the bi-partition labeled by $g$ and $\xi$ (see Table~\ref{table:honeycomb}).
Note that here we set $\lambda = 0$ in Eq.~(\ref{eq:L}), meaning different values of the LN are allowed for different bi-partition $\xi$ of the same group $g$. 
Figure~\ref{Fig:appendix_logneg_con}(a) shows one of the optimized interactions with the smallest $L (=-\bar{E}^{\mathcal{N}}) \sim -1.78$ among 30 calculations for the honeycomb lattice model. 
Six out of nine interactions are completely anisotropic, similar to the Kitaev model, while the remaining three take more isotropic values. 
Thus, this optimization procedure finds a model with a mixture of the Kitaev-type anisotropic interactions and the Heisenberg-like isotropic interactions. 

Next, we perform calculations based on the parametrization in Eq.~(\ref{eq:Jij_2}), keeping the other setup the same as above. 
Figure~\ref{Fig:appendix_logneg_con}(b) shows the optimized interactions. 
Remarkably, in this case all the interactions are strongly anisotropic, showing different amplitudes of the interactions depending on the bonds. 
The amplitudes are represented by the thickness of each bond proportional to $\sqrt{\sum_\mu (J_{ij}^{\mu})^2}$. 
The optimized values are $|J_{14}^x|, |J_{23}^x|, |J_{56}^x| \simeq 0.517$ and $|J_{16}^y|, |J_{25}^y|, |J_{34}^y|, |J_{12}^z|, |J_{36}^z|, |J_{45}^z| \simeq 1.168$, and the other components of $J_{ij}^\mu$ are almost zero. 
This configuration corresponds to a spatially-anisotropic Kitaev model in which the $J_x$ bonds are weaker than $J_y$ and $J_z$. 
Interestingly, the value of $L$ for this solution is $L (=-\bar{E}^{\mathcal{N}}) \sim -1.80$, which is smaller than the result with the parametrization using Eq.~(\ref{eq:Jij}).

This comparison demonstrates that better parametrization depends on the problem. 
Choices of the objective function and the model parametrization generate different Hamiltonians. 
Therefore, one should choose how to parameterize the interaction carefully according to the purpose. 


\section{Computational instability}
\label{sec:app_instability}

\begin{figure}[htbp]
        \centering
        \includegraphics[width=1.0\columnwidth,pagebox=cropbox,clip]{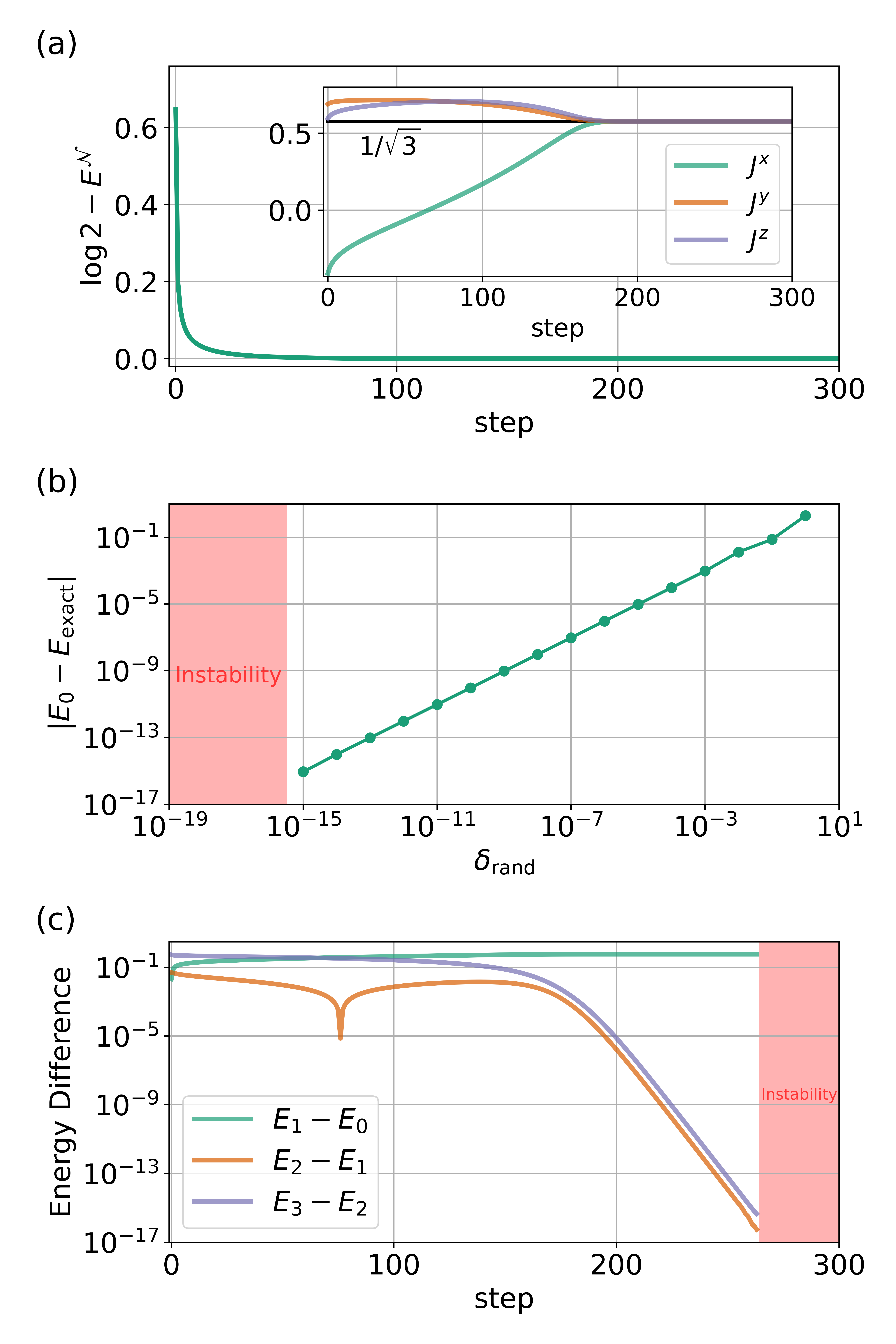}
        \caption{
        (a) Changes in the logarithmic negativity $E^{\mathcal{N}}$ through the optimization process for the two-spin model. 
        The inset shows changes in each component of spin interactions, $J^{\mu}$. 
        We set $\eta=0.03$ during the optimization. 
        (b) Deviation of the ground state energy of the optimized result, $E_0$, from the exact energy for $J^x=J^y=J^z=1/\sqrt{3}$, $E_{\rm exact} = - \sqrt{3}/4$, as a function of the variance of the random matrix elements, $\delta_{\rm rand}$. 
        (c) Changes in the energy difference between the low-energy eigenstates. 
        $E_i$ is the energy of the $i$th excited state. 
        In (b) and (c), the red area represents the region where the calculations become unstable. 
}
        \label{Fig:appendix_2site}
\end{figure}

In this Appendix, we investigate the computational instability regarding the automatic differentiation of diagonalization. 
The JAX library does not currently support the automatic differentiation of diagonalization when involving degeneracy in the eigenvalues. 
To circumvent this issue, we add a tiny random Hermitian matrix, denoted as $\delta \mathcal{H}_{\rm rand}$, to the Hamiltonian.
This minor modification slightly lift the degeneracy. 
Each element of $\delta \mathcal{H}_{\rm rand}$ is generated from a Gaussian distribution with a mean of $0$ and a variance of $\delta_{\rm rand}$. 
In the optimization process, we use a different random matrix at each step, thereby facilitating more robust calculations.

Let us demonstrate the instability using a two-spin model as an example.
The Hamiltonian is defined by $\mathcal{H} = J^x \hat{\sigma}^x_1 \hat{\sigma}^x_2 + J^y \hat{\sigma}^y_1 \hat{\sigma}^y_2 +  J^z \hat{\sigma}^z_1 \hat{\sigma}^z_2 $ with the condition $({J^x})^2+ ({J^y})^2+ ({J^z})^2 = 1$. 
Here, we use the LN discussed in Appendix \ref{sec:mutinfo} as the objective function, since the TEEE for the two-spin model takes almost always the maximum value for any interaction.
Note that, however, the following arguments are applicable to any objective functions that requires the numerical diagonalization.

Figure~\ref{Fig:appendix_2site}(a) shows the optimization process of the objective function $L=-E_{\mathcal{N}}$, with the interaction parameters $\{ J^x, J^y, J^z \}$ in the inset. 
We set $\beta=20$ and $\delta_{\rm rand} = 10^{-10}$.
Note that $\bar{E}^{\mathcal{N}} = E^{\mathcal{N}}$ since there is only one way for bi-partition in the two-spin problem. 
In this case, the LN converges to $\log2$, and the interaction parameters converge $J^x=J^y=J^z=\frac{1}{\sqrt{3}}$; 
namely, the optimization leads to the AFM Heisenberg model with isotropic interactions.
In Fig.~\ref{Fig:appendix_2site}(b), we plot the deviation of the ground state energy, $E_0$, from that of AFM Heisenberg model, $E_{\rm exact}$, while changing $\delta_{\rm rand}$. 
It clearly shows that the deviation is proportional to $\delta_{\rm rand}$,  
while the calculations for $\delta_{\rm rand} < 10^{-15}$ becomes unstable as indicated by red shade. 
Figure~\ref{Fig:appendix_2site}(c) shows the changes in the energy differences between the adjacent states when $\delta_{\rm rand} =0$; 
here $E_i$ is the energy of the $i$th excited state.  
It shows that computational instability arises when the energy differences, for instance, $E_3-E_2$ and $E_2-E_1$, are reduced below $\sim 10^{-16}$. 
This indicates that the optimization breaks down without $\delta \mathcal{H}_{\rm rand}$ due to the degeneracy. 
For the stable computation, we take $\delta_{\rm rand}$ in the order of $10^{-8}$ in the calculations in the main text as well as other Appendices.

\section{Model purification}
\label{sec:S_params}

\begin{figure}[htbp]
        \centering
        \includegraphics[width=0.95\columnwidth,pagebox=cropbox,trim={0 0 0 0},clip]{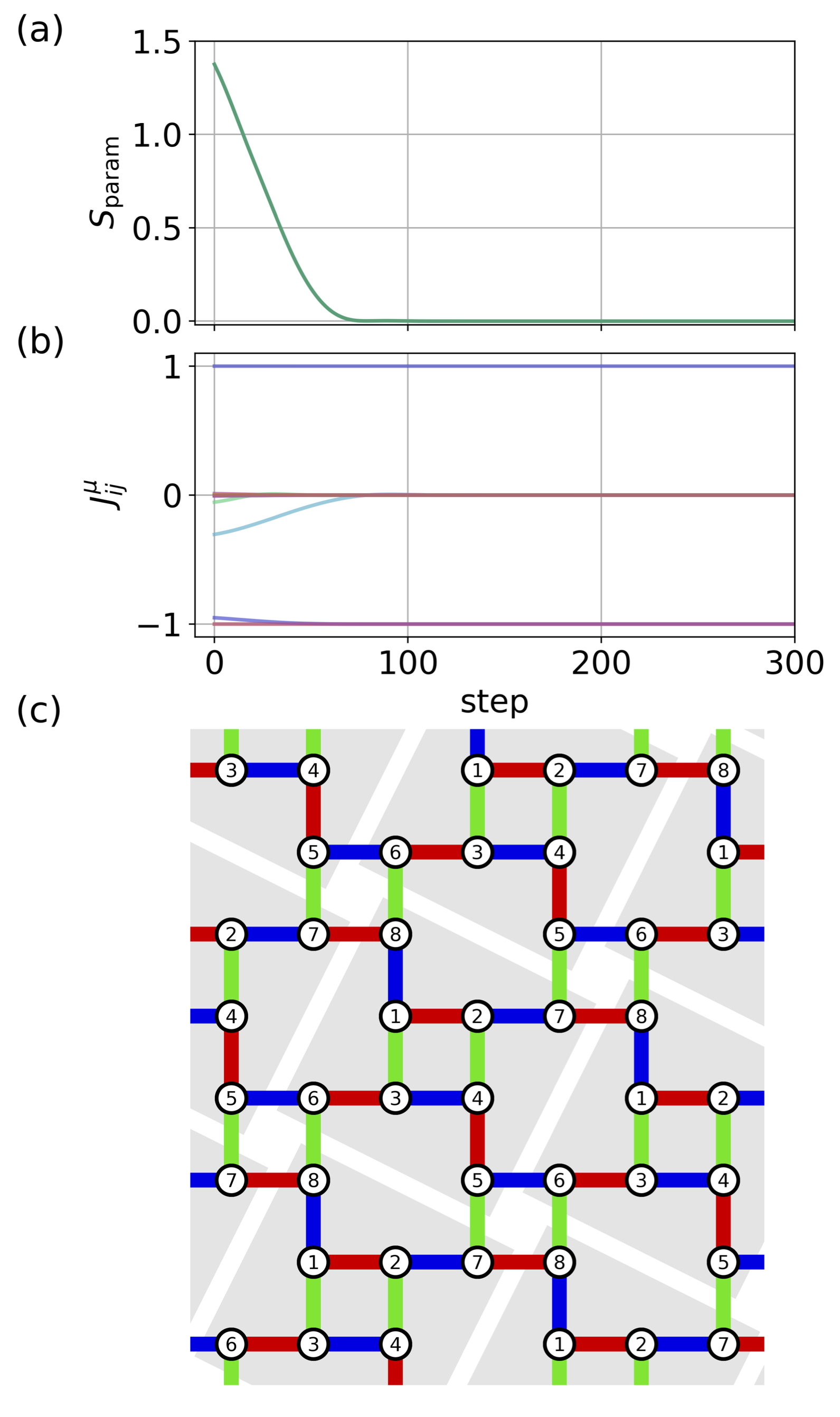}
        \caption{
        (a) Changes in $S_{\rm param} $ through the optimization process using Eq.~(\ref{eq:L_param}) starting from the result in Fig.~\ref{Fig:48}(a) for the model on the square-octagon lattice.  
        We set $\tilde{\lambda}=1000$ and $\eta=0.01$ during the optimization. 
        (b) Changes in the interaction parameters. 
        (c) Optimized interaction parameters. 
        The notations are common to those in Fig.~\ref{Fig:48}. 
}
        \label{Fig:appendix_48}
\end{figure}

\begin{figure}[htbp]
        \centering
        \includegraphics[width=0.95\columnwidth,pagebox=cropbox,trim={0 0 0 0},clip]{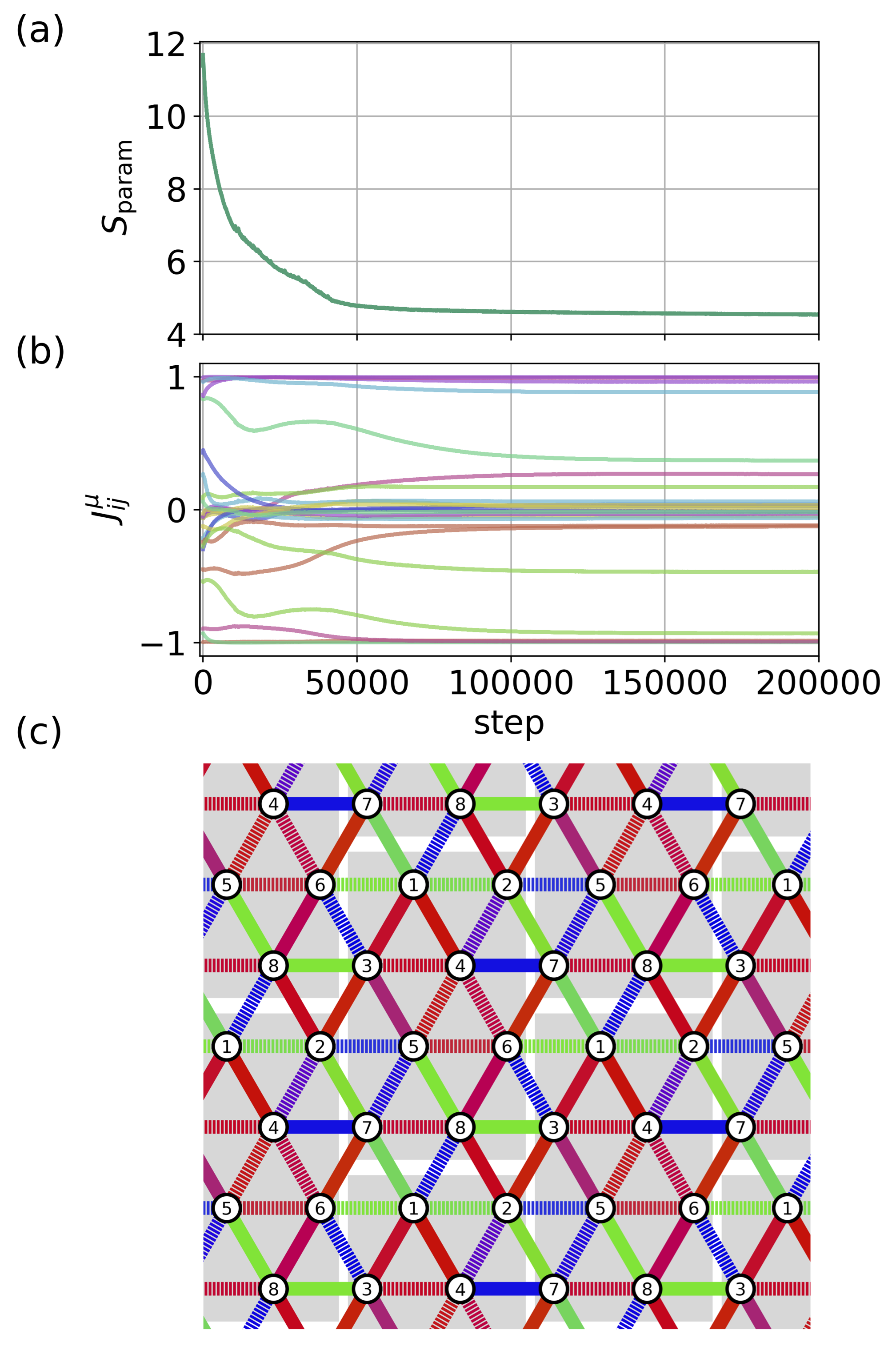}
        \caption{
        (a) Changes in $S_{\rm param} $ through the optimization process using Eq.~(\ref{eq:L_param}) starting from the result in Fig.~\ref{Fig:triangular} for the model on the triangular lattice.  
        We set $\tilde{\lambda}=1000$ and $\eta=0.03$ during the optimization. 
        (b) Changes in the interaction parameters. 
        (c) Optimized interaction parameters. 
        The notations are common to those in Fig.~\ref{Fig:triangular}. 
}
        \label{Fig:appendix_tri}
\end{figure}

In this Appendix, we propose a method to purify the model obtained by the optimization process. 
In general, the optimization of maximizing entanglement results in a model with spatially inhomogeneous interaction parameters. 
In such cases, it is insightful if it is able to transform the results into a more simple model without changing the value of the entanglement.
To achieve this, we introduce the entropy associated with the simpleness of the interaction parameters, which we call the parameter entropy as 
\begin{align}
S_{\rm param} = \sum_{ij\mu} p_{ij}^\mu \log p_{ij}^\mu \label{S_para},
\end{align}
where 
\begin{align}
p_{ij}^\mu = \frac{|J_{ij}^{\mu}|^2}{\sum_{ij\mu} |J_{ij}^{\mu}|^2}.  
\end{align}
The smaller the parameter entropy, the fewer nonzero components in the interaction parameters. 
We optimize the parameter entropy to be smaller after the optimization of $L$ using TEEE; 
namely, we perform an additional optimization using the following objective function, 
\begin{align}
\tilde{L}({\bm \theta }) = S_{\rm param}({\bm \theta }) + \tilde{\lambda} \Big({\rm ReLU}\big(L_{\rm opt}-L({\bm \theta })\big)\Big)^2, 
\label{eq:L_param}
\end{align}
where $\tilde{\lambda}$ is the hyperparameter, $\rm ReLU$ denotes the rectified linear unit function, $L_{\rm opt}$ is the final value $L$ after the optimization of the TEEE, and $L({\bm \theta})$ in the input for ReLU is the same as Eq.~(\ref{eq:L}). 
By optimizing $\bm \theta$ to minimize the objective function $\tilde{L}({\bm \theta })$, we can obtain a simpler model without increasing the value of $L$.

First, we apply this purification to the results of maximizing the TEEE on the square-octagon lattice in Sec.~\ref{subsec:48}. 
The interactions in Fig.~\ref{Fig:48}(a) are not completely homogeneous, with some bonds exhibiting slight deviations from the ideal Kitaev model.
Figure~\ref{Fig:appendix_48}(a) shows the optimization process of the parameter entropy.
It is successfully reduced to zero, indicating that each bond has only one nonzero component among $\{J^x, J^y, J^z\}$ as shown in Fig.~\ref{Fig:appendix_48}(b). 
Figure~\ref{Fig:appendix_48}(c) presents the interactions after the purification. 
Compared to Fig.~\ref{Fig:48}(a), all bonds become completely anisotropic as in the ideal Kitaev model. 
This comparison confirms that minimizing the parameter entropy is useful to purify the models, yielding more interpretable and tractable models.



Next, we apply the method to the results of the triangular lattice in Sec.~\ref{subsec:tri}. 
As indicated in Fig.~\ref{Fig:appendix_tri}(a), the optimization process starting from the result in Fig.~\ref{Fig:triangular} leads to a reduction in the value of $S_{\rm param}$, although it does not achieve a value of zero, which is different from the square-octagon lattice case. 
The optimization process for $J_{ij}^\mu$ in Fig.~\ref{Fig:appendix_tri}(b) shows that the interactions are not homogeneous,  while there is a change in the values of $J_{ij}^\mu$. 
The interactions after the purification lack any clear pattern as illustrated in Fig.~\ref{Fig:appendix_tri}(c). 
These results indicate that this purification method using the parameter entropy does not always successfully reduce the inhomogeneity of the model. 
The development of alternative objective functions for the purification may yield more robust results; it will be a focus for future research.



\bibliography{main}

\begin{thebibliography}{64}%
\makeatletter
\providecommand \@ifxundefined [1]{%
 \@ifx{#1\undefined}
}%
\providecommand \@ifnum [1]{%
 \ifnum #1\expandafter \@firstoftwo
 \else \expandafter \@secondoftwo
 \fi
}%
\providecommand \@ifx [1]{%
 \ifx #1\expandafter \@firstoftwo
 \else \expandafter \@secondoftwo
 \fi
}%
\providecommand \natexlab [1]{#1}%
\providecommand \enquote  [1]{``#1''}%
\providecommand \bibnamefont  [1]{#1}%
\providecommand \bibfnamefont [1]{#1}%
\providecommand \citenamefont [1]{#1}%
\providecommand \href@noop [0]{\@secondoftwo}%
\providecommand \href [0]{\begingroup \@sanitize@url \@href}%
\providecommand \@href[1]{\@@startlink{#1}\@@href}%
\providecommand \@@href[1]{\endgroup#1\@@endlink}%
\providecommand \@sanitize@url [0]{\catcode `\\12\catcode `\$12\catcode
  `\&12\catcode `\#12\catcode `\^12\catcode `\_12\catcode `\%12\relax}%
\providecommand \@@startlink[1]{}%
\providecommand \@@endlink[0]{}%
\providecommand \url  [0]{\begingroup\@sanitize@url \@url }%
\providecommand \@url [1]{\endgroup\@href {#1}{\urlprefix }}%
\providecommand \urlprefix  [0]{URL }%
\providecommand \Eprint [0]{\href }%
\providecommand \doibase [0]{https://doi.org/}%
\providecommand \selectlanguage [0]{\@gobble}%
\providecommand \bibinfo  [0]{\@secondoftwo}%
\providecommand \bibfield  [0]{\@secondoftwo}%
\providecommand \translation [1]{[#1]}%
\providecommand \BibitemOpen [0]{}%
\providecommand \bibitemStop [0]{}%
\providecommand \bibitemNoStop [0]{.\EOS\space}%
\providecommand \EOS [0]{\spacefactor3000\relax}%
\providecommand \BibitemShut  [1]{\csname bibitem#1\endcsname}%
\let\auto@bib@innerbib\@empty
\bibitem [{\citenamefont {Almheiri}\ \emph {et~al.}(2021)\citenamefont
  {Almheiri}, \citenamefont {Hartman}, \citenamefont {Maldacena}, \citenamefont
  {Shaghoulian},\ and\ \citenamefont {Tajdini}}]{RevModPhys.93.035002}%
  \BibitemOpen
  \bibfield  {author} {\bibinfo {author} {\bibfnamefont {A.}~\bibnamefont
  {Almheiri}}, \bibinfo {author} {\bibfnamefont {T.}~\bibnamefont {Hartman}},
  \bibinfo {author} {\bibfnamefont {J.}~\bibnamefont {Maldacena}}, \bibinfo
  {author} {\bibfnamefont {E.}~\bibnamefont {Shaghoulian}},\ and\ \bibinfo
  {author} {\bibfnamefont {A.}~\bibnamefont {Tajdini}},\ }\bibfield  {title}
  {\bibinfo {title} {The entropy of hawking radiation},\ }\href
  {https://doi.org/10.1103/RevModPhys.93.035002} {\bibfield  {journal}
  {\bibinfo  {journal} {Rev. Mod. Phys.}\ }\textbf {\bibinfo {volume} {93}},\
  \bibinfo {pages} {035002} (\bibinfo {year} {2021})}\BibitemShut {NoStop}%
\bibitem [{\citenamefont {Amico}\ \emph {et~al.}(2008)\citenamefont {Amico},
  \citenamefont {Fazio}, \citenamefont {Osterloh},\ and\ \citenamefont
  {Vedral}}]{RevModPhys.80.517}%
  \BibitemOpen
  \bibfield  {author} {\bibinfo {author} {\bibfnamefont {L.}~\bibnamefont
  {Amico}}, \bibinfo {author} {\bibfnamefont {R.}~\bibnamefont {Fazio}},
  \bibinfo {author} {\bibfnamefont {A.}~\bibnamefont {Osterloh}},\ and\
  \bibinfo {author} {\bibfnamefont {V.}~\bibnamefont {Vedral}},\ }\bibfield
  {title} {\bibinfo {title} {Entanglement in many-body systems},\ }\href
  {https://doi.org/10.1103/RevModPhys.80.517} {\bibfield  {journal} {\bibinfo
  {journal} {Rev. Mod. Phys.}\ }\textbf {\bibinfo {volume} {80}},\ \bibinfo
  {pages} {517} (\bibinfo {year} {2008})}\BibitemShut {NoStop}%
\bibitem [{\citenamefont {Bennett}\ \emph
  {et~al.}(1996{\natexlab{a}})\citenamefont {Bennett}, \citenamefont
  {DiVincenzo}, \citenamefont {Smolin},\ and\ \citenamefont
  {Wootters}}]{PhysRevA.54.3824}%
  \BibitemOpen
  \bibfield  {author} {\bibinfo {author} {\bibfnamefont {C.~H.}\ \bibnamefont
  {Bennett}}, \bibinfo {author} {\bibfnamefont {D.~P.}\ \bibnamefont
  {DiVincenzo}}, \bibinfo {author} {\bibfnamefont {J.~A.}\ \bibnamefont
  {Smolin}},\ and\ \bibinfo {author} {\bibfnamefont {W.~K.}\ \bibnamefont
  {Wootters}},\ }\bibfield  {title} {\bibinfo {title} {Mixed-state entanglement
  and quantum error correction},\ }\href
  {https://doi.org/10.1103/PhysRevA.54.3824} {\bibfield  {journal} {\bibinfo
  {journal} {Phys. Rev. A}\ }\textbf {\bibinfo {volume} {54}},\ \bibinfo
  {pages} {3824} (\bibinfo {year} {1996}{\natexlab{a}})}\BibitemShut {NoStop}%
\bibitem [{\citenamefont {Van~den Nest}\ \emph {et~al.}(2006)\citenamefont
  {Van~den Nest}, \citenamefont {Miyake}, \citenamefont {D\"ur},\ and\
  \citenamefont {Briegel}}]{PhysRevLett.97.150504}%
  \BibitemOpen
  \bibfield  {author} {\bibinfo {author} {\bibfnamefont {M.}~\bibnamefont
  {Van~den Nest}}, \bibinfo {author} {\bibfnamefont {A.}~\bibnamefont
  {Miyake}}, \bibinfo {author} {\bibfnamefont {W.}~\bibnamefont {D\"ur}},\ and\
  \bibinfo {author} {\bibfnamefont {H.~J.}\ \bibnamefont {Briegel}},\
  }\bibfield  {title} {\bibinfo {title} {Universal resources for
  measurement-based quantum computation},\ }\href
  {https://doi.org/10.1103/PhysRevLett.97.150504} {\bibfield  {journal}
  {\bibinfo  {journal} {Phys. Rev. Lett.}\ }\textbf {\bibinfo {volume} {97}},\
  \bibinfo {pages} {150504} (\bibinfo {year} {2006})}\BibitemShut {NoStop}%
\bibitem [{\citenamefont {Gross}\ \emph {et~al.}(2009)\citenamefont {Gross},
  \citenamefont {Flammia},\ and\ \citenamefont
  {Eisert}}]{PhysRevLett.102.190501}%
  \BibitemOpen
  \bibfield  {author} {\bibinfo {author} {\bibfnamefont {D.}~\bibnamefont
  {Gross}}, \bibinfo {author} {\bibfnamefont {S.~T.}\ \bibnamefont {Flammia}},\
  and\ \bibinfo {author} {\bibfnamefont {J.}~\bibnamefont {Eisert}},\
  }\bibfield  {title} {\bibinfo {title} {Most quantum states are too entangled
  to be useful as computational resources},\ }\href
  {https://doi.org/10.1103/PhysRevLett.102.190501} {\bibfield  {journal}
  {\bibinfo  {journal} {Phys. Rev. Lett.}\ }\textbf {\bibinfo {volume} {102}},\
  \bibinfo {pages} {190501} (\bibinfo {year} {2009})}\BibitemShut {NoStop}%
\bibitem [{\citenamefont {Choi}\ \emph {et~al.}(2020)\citenamefont {Choi},
  \citenamefont {Bao}, \citenamefont {Qi},\ and\ \citenamefont
  {Altman}}]{PhysRevLett.125.030505}%
  \BibitemOpen
  \bibfield  {author} {\bibinfo {author} {\bibfnamefont {S.}~\bibnamefont
  {Choi}}, \bibinfo {author} {\bibfnamefont {Y.}~\bibnamefont {Bao}}, \bibinfo
  {author} {\bibfnamefont {X.-L.}\ \bibnamefont {Qi}},\ and\ \bibinfo {author}
  {\bibfnamefont {E.}~\bibnamefont {Altman}},\ }\bibfield  {title} {\bibinfo
  {title} {Quantum error correction in scrambling dynamics and
  measurement-induced phase transition},\ }\href
  {https://doi.org/10.1103/PhysRevLett.125.030505} {\bibfield  {journal}
  {\bibinfo  {journal} {Phys. Rev. Lett.}\ }\textbf {\bibinfo {volume} {125}},\
  \bibinfo {pages} {030505} (\bibinfo {year} {2020})}\BibitemShut {NoStop}%
\bibitem [{\citenamefont {Li}\ and\ \citenamefont
  {Fisher}(2021)}]{PhysRevB.103.104306}%
  \BibitemOpen
  \bibfield  {author} {\bibinfo {author} {\bibfnamefont {Y.}~\bibnamefont
  {Li}}\ and\ \bibinfo {author} {\bibfnamefont {M.~P.~A.}\ \bibnamefont
  {Fisher}},\ }\bibfield  {title} {\bibinfo {title} {Statistical mechanics of
  quantum error correcting codes},\ }\href
  {https://doi.org/10.1103/PhysRevB.103.104306} {\bibfield  {journal} {\bibinfo
   {journal} {Phys. Rev. B}\ }\textbf {\bibinfo {volume} {103}},\ \bibinfo
  {pages} {104306} (\bibinfo {year} {2021})}\BibitemShut {NoStop}%
\bibitem [{\citenamefont {Castelnovo}\ and\ \citenamefont
  {Chamon}(2007)}]{PhysRevB.76.184442}%
  \BibitemOpen
  \bibfield  {author} {\bibinfo {author} {\bibfnamefont {C.}~\bibnamefont
  {Castelnovo}}\ and\ \bibinfo {author} {\bibfnamefont {C.}~\bibnamefont
  {Chamon}},\ }\bibfield  {title} {\bibinfo {title} {Entanglement and
  topological entropy of the toric code at finite temperature},\ }\href
  {https://doi.org/10.1103/PhysRevB.76.184442} {\bibfield  {journal} {\bibinfo
  {journal} {Phys. Rev. B}\ }\textbf {\bibinfo {volume} {76}},\ \bibinfo
  {pages} {184442} (\bibinfo {year} {2007})}\BibitemShut {NoStop}%
\bibitem [{\citenamefont {Ma}\ \emph {et~al.}(2018)\citenamefont {Ma},
  \citenamefont {Schmitz}, \citenamefont {Parameswaran}, \citenamefont
  {Hermele},\ and\ \citenamefont {Nandkishore}}]{PhysRevB.97.125101}%
  \BibitemOpen
  \bibfield  {author} {\bibinfo {author} {\bibfnamefont {H.}~\bibnamefont
  {Ma}}, \bibinfo {author} {\bibfnamefont {A.~T.}\ \bibnamefont {Schmitz}},
  \bibinfo {author} {\bibfnamefont {S.~A.}\ \bibnamefont {Parameswaran}},
  \bibinfo {author} {\bibfnamefont {M.}~\bibnamefont {Hermele}},\ and\ \bibinfo
  {author} {\bibfnamefont {R.~M.}\ \bibnamefont {Nandkishore}},\ }\bibfield
  {title} {\bibinfo {title} {Topological entanglement entropy of fracton
  stabilizer codes},\ }\href {https://doi.org/10.1103/PhysRevB.97.125101}
  {\bibfield  {journal} {\bibinfo  {journal} {Phys. Rev. B}\ }\textbf {\bibinfo
  {volume} {97}},\ \bibinfo {pages} {125101} (\bibinfo {year}
  {2018})}\BibitemShut {NoStop}%
\bibitem [{\citenamefont {Grover}\ \emph {et~al.}(2011)\citenamefont {Grover},
  \citenamefont {Turner},\ and\ \citenamefont
  {Vishwanath}}]{PhysRevB.84.195120}%
  \BibitemOpen
  \bibfield  {author} {\bibinfo {author} {\bibfnamefont {T.}~\bibnamefont
  {Grover}}, \bibinfo {author} {\bibfnamefont {A.~M.}\ \bibnamefont {Turner}},\
  and\ \bibinfo {author} {\bibfnamefont {A.}~\bibnamefont {Vishwanath}},\
  }\bibfield  {title} {\bibinfo {title} {Entanglement entropy of gapped phases
  and topological order in three dimensions},\ }\href
  {https://doi.org/10.1103/PhysRevB.84.195120} {\bibfield  {journal} {\bibinfo
  {journal} {Phys. Rev. B}\ }\textbf {\bibinfo {volume} {84}},\ \bibinfo
  {pages} {195120} (\bibinfo {year} {2011})}\BibitemShut {NoStop}%
\bibitem [{\citenamefont {Vidal}\ \emph {et~al.}(2003)\citenamefont {Vidal},
  \citenamefont {Latorre}, \citenamefont {Rico},\ and\ \citenamefont
  {Kitaev}}]{PhysRevLett.90.227902}%
  \BibitemOpen
  \bibfield  {author} {\bibinfo {author} {\bibfnamefont {G.}~\bibnamefont
  {Vidal}}, \bibinfo {author} {\bibfnamefont {J.~I.}\ \bibnamefont {Latorre}},
  \bibinfo {author} {\bibfnamefont {E.}~\bibnamefont {Rico}},\ and\ \bibinfo
  {author} {\bibfnamefont {A.}~\bibnamefont {Kitaev}},\ }\bibfield  {title}
  {\bibinfo {title} {Entanglement in quantum critical phenomena},\ }\href
  {https://doi.org/10.1103/PhysRevLett.90.227902} {\bibfield  {journal}
  {\bibinfo  {journal} {Phys. Rev. Lett.}\ }\textbf {\bibinfo {volume} {90}},\
  \bibinfo {pages} {227902} (\bibinfo {year} {2003})}\BibitemShut {NoStop}%
\bibitem [{\citenamefont {Osborne}\ and\ \citenamefont
  {Nielsen}(2002)}]{PhysRevA.66.032110}%
  \BibitemOpen
  \bibfield  {author} {\bibinfo {author} {\bibfnamefont {T.~J.}\ \bibnamefont
  {Osborne}}\ and\ \bibinfo {author} {\bibfnamefont {M.~A.}\ \bibnamefont
  {Nielsen}},\ }\bibfield  {title} {\bibinfo {title} {Entanglement in a simple
  quantum phase transition},\ }\href
  {https://doi.org/10.1103/PhysRevA.66.032110} {\bibfield  {journal} {\bibinfo
  {journal} {Phys. Rev. A}\ }\textbf {\bibinfo {volume} {66}},\ \bibinfo
  {pages} {032110} (\bibinfo {year} {2002})}\BibitemShut {NoStop}%
\bibitem [{\citenamefont {Osterloh}\ \emph {et~al.}(2002)\citenamefont
  {Osterloh}, \citenamefont {Amico}, \citenamefont {Falci},\ and\ \citenamefont
  {Fazio}}]{Osterloh2002}%
  \BibitemOpen
  \bibfield  {author} {\bibinfo {author} {\bibfnamefont {A.}~\bibnamefont
  {Osterloh}}, \bibinfo {author} {\bibfnamefont {L.}~\bibnamefont {Amico}},
  \bibinfo {author} {\bibfnamefont {G.}~\bibnamefont {Falci}},\ and\ \bibinfo
  {author} {\bibfnamefont {R.}~\bibnamefont {Fazio}},\ }\bibfield  {title}
  {\bibinfo {title} {Scaling of entanglement close to a quantum phase
  transition},\ }\href {https://doi.org/10.1038/416608a} {\bibfield  {journal}
  {\bibinfo  {journal} {Nature}\ }\textbf {\bibinfo {volume} {416}},\ \bibinfo
  {pages} {608} (\bibinfo {year} {2002})}\BibitemShut {NoStop}%
\bibitem [{\citenamefont {Shen}\ \emph {et~al.}(2017)\citenamefont {Shen},
  \citenamefont {Zhang}, \citenamefont {Fan},\ and\ \citenamefont
  {Zhai}}]{PhysRevB.96.054503}%
  \BibitemOpen
  \bibfield  {author} {\bibinfo {author} {\bibfnamefont {H.}~\bibnamefont
  {Shen}}, \bibinfo {author} {\bibfnamefont {P.}~\bibnamefont {Zhang}},
  \bibinfo {author} {\bibfnamefont {R.}~\bibnamefont {Fan}},\ and\ \bibinfo
  {author} {\bibfnamefont {H.}~\bibnamefont {Zhai}},\ }\bibfield  {title}
  {\bibinfo {title} {Out-of-time-order correlation at a quantum phase
  transition},\ }\href {https://doi.org/10.1103/PhysRevB.96.054503} {\bibfield
  {journal} {\bibinfo  {journal} {Phys. Rev. B}\ }\textbf {\bibinfo {volume}
  {96}},\ \bibinfo {pages} {054503} (\bibinfo {year} {2017})}\BibitemShut
  {NoStop}%
\bibitem [{\citenamefont {Dillenschneider}(2008)}]{PhysRevB.78.224413}%
  \BibitemOpen
  \bibfield  {author} {\bibinfo {author} {\bibfnamefont {R.}~\bibnamefont
  {Dillenschneider}},\ }\bibfield  {title} {\bibinfo {title} {Quantum discord
  and quantum phase transition in spin chains},\ }\href
  {https://doi.org/10.1103/PhysRevB.78.224413} {\bibfield  {journal} {\bibinfo
  {journal} {Phys. Rev. B}\ }\textbf {\bibinfo {volume} {78}},\ \bibinfo
  {pages} {224413} (\bibinfo {year} {2008})}\BibitemShut {NoStop}%
\bibitem [{\citenamefont {Canovi}\ \emph {et~al.}(2014)\citenamefont {Canovi},
  \citenamefont {Ercolessi}, \citenamefont {Naldesi}, \citenamefont {Taddia},\
  and\ \citenamefont {Vodola}}]{PhysRevB.89.104303}%
  \BibitemOpen
  \bibfield  {author} {\bibinfo {author} {\bibfnamefont {E.}~\bibnamefont
  {Canovi}}, \bibinfo {author} {\bibfnamefont {E.}~\bibnamefont {Ercolessi}},
  \bibinfo {author} {\bibfnamefont {P.}~\bibnamefont {Naldesi}}, \bibinfo
  {author} {\bibfnamefont {L.}~\bibnamefont {Taddia}},\ and\ \bibinfo {author}
  {\bibfnamefont {D.}~\bibnamefont {Vodola}},\ }\bibfield  {title} {\bibinfo
  {title} {Dynamics of entanglement entropy and entanglement spectrum crossing
  a quantum phase transition},\ }\href
  {https://doi.org/10.1103/PhysRevB.89.104303} {\bibfield  {journal} {\bibinfo
  {journal} {Phys. Rev. B}\ }\textbf {\bibinfo {volume} {89}},\ \bibinfo
  {pages} {104303} (\bibinfo {year} {2014})}\BibitemShut {NoStop}%
\bibitem [{\citenamefont {Ling}\ \emph {et~al.}(2016)\citenamefont {Ling},
  \citenamefont {Liu},\ and\ \citenamefont {Wu}}]{PhysRevD.93.126004}%
  \BibitemOpen
  \bibfield  {author} {\bibinfo {author} {\bibfnamefont {Y.}~\bibnamefont
  {Ling}}, \bibinfo {author} {\bibfnamefont {P.}~\bibnamefont {Liu}},\ and\
  \bibinfo {author} {\bibfnamefont {J.-P.}\ \bibnamefont {Wu}},\ }\bibfield
  {title} {\bibinfo {title} {Characterization of quantum phase transition using
  holographic entanglement entropy},\ }\href
  {https://doi.org/10.1103/PhysRevD.93.126004} {\bibfield  {journal} {\bibinfo
  {journal} {Phys. Rev. D}\ }\textbf {\bibinfo {volume} {93}},\ \bibinfo
  {pages} {126004} (\bibinfo {year} {2016})}\BibitemShut {NoStop}%
\bibitem [{\citenamefont {Valdez}\ \emph {et~al.}(2017)\citenamefont {Valdez},
  \citenamefont {Jaschke}, \citenamefont {Vargas},\ and\ \citenamefont
  {Carr}}]{PhysRevLett.119.225301}%
  \BibitemOpen
  \bibfield  {author} {\bibinfo {author} {\bibfnamefont {M.~A.}\ \bibnamefont
  {Valdez}}, \bibinfo {author} {\bibfnamefont {D.}~\bibnamefont {Jaschke}},
  \bibinfo {author} {\bibfnamefont {D.~L.}\ \bibnamefont {Vargas}},\ and\
  \bibinfo {author} {\bibfnamefont {L.~D.}\ \bibnamefont {Carr}},\ }\bibfield
  {title} {\bibinfo {title} {Quantifying complexity in quantum phase
  transitions via mutual information complex networks},\ }\href
  {https://doi.org/10.1103/PhysRevLett.119.225301} {\bibfield  {journal}
  {\bibinfo  {journal} {Phys. Rev. Lett.}\ }\textbf {\bibinfo {volume} {119}},\
  \bibinfo {pages} {225301} (\bibinfo {year} {2017})}\BibitemShut {NoStop}%
\bibitem [{\citenamefont {Kitaev}\ and\ \citenamefont
  {Preskill}(2006)}]{PhysRevLett.96.110404}%
  \BibitemOpen
  \bibfield  {author} {\bibinfo {author} {\bibfnamefont {A.}~\bibnamefont
  {Kitaev}}\ and\ \bibinfo {author} {\bibfnamefont {J.}~\bibnamefont
  {Preskill}},\ }\bibfield  {title} {\bibinfo {title} {Topological entanglement
  entropy},\ }\href {https://doi.org/10.1103/PhysRevLett.96.110404} {\bibfield
  {journal} {\bibinfo  {journal} {Phys. Rev. Lett.}\ }\textbf {\bibinfo
  {volume} {96}},\ \bibinfo {pages} {110404} (\bibinfo {year}
  {2006})}\BibitemShut {NoStop}%
\bibitem [{\citenamefont {Levin}\ and\ \citenamefont
  {Wen}(2006)}]{PhysRevLett.96.110405}%
  \BibitemOpen
  \bibfield  {author} {\bibinfo {author} {\bibfnamefont {M.}~\bibnamefont
  {Levin}}\ and\ \bibinfo {author} {\bibfnamefont {X.-G.}\ \bibnamefont
  {Wen}},\ }\bibfield  {title} {\bibinfo {title} {Detecting topological order
  in a ground state wave function},\ }\href
  {https://doi.org/10.1103/PhysRevLett.96.110405} {\bibfield  {journal}
  {\bibinfo  {journal} {Phys. Rev. Lett.}\ }\textbf {\bibinfo {volume} {96}},\
  \bibinfo {pages} {110405} (\bibinfo {year} {2006})}\BibitemShut {NoStop}%
\bibitem [{\citenamefont {Jiang}\ \emph {et~al.}(2012)\citenamefont {Jiang},
  \citenamefont {Wang},\ and\ \citenamefont {Balents}}]{Jiang2012}%
  \BibitemOpen
  \bibfield  {author} {\bibinfo {author} {\bibfnamefont {H.-C.}\ \bibnamefont
  {Jiang}}, \bibinfo {author} {\bibfnamefont {Z.}~\bibnamefont {Wang}},\ and\
  \bibinfo {author} {\bibfnamefont {L.}~\bibnamefont {Balents}},\ }\bibfield
  {title} {\bibinfo {title} {Identifying topological order by entanglement
  entropy},\ }\href {https://doi.org/10.1038/nphys2465} {\bibfield  {journal}
  {\bibinfo  {journal} {Nature Physics}\ }\textbf {\bibinfo {volume} {8}},\
  \bibinfo {pages} {902} (\bibinfo {year} {2012})}\BibitemShut {NoStop}%
\bibitem [{\citenamefont {Isakov}\ \emph {et~al.}(2011)\citenamefont {Isakov},
  \citenamefont {Hastings},\ and\ \citenamefont {Melko}}]{Isakov2011}%
  \BibitemOpen
  \bibfield  {author} {\bibinfo {author} {\bibfnamefont {S.~V.}\ \bibnamefont
  {Isakov}}, \bibinfo {author} {\bibfnamefont {M.~B.}\ \bibnamefont
  {Hastings}},\ and\ \bibinfo {author} {\bibfnamefont {R.~G.}\ \bibnamefont
  {Melko}},\ }\bibfield  {title} {\bibinfo {title} {Topological entanglement
  entropy of a bose--hubbard spin liquid},\ }\href
  {https://doi.org/10.1038/nphys2036} {\bibfield  {journal} {\bibinfo
  {journal} {Nature Physics}\ }\textbf {\bibinfo {volume} {7}},\ \bibinfo
  {pages} {772} (\bibinfo {year} {2011})}\BibitemShut {NoStop}%
\bibitem [{\citenamefont {Solodukhin}(2011)}]{Solodukhin2011}%
  \BibitemOpen
  \bibfield  {author} {\bibinfo {author} {\bibfnamefont {S.~N.}\ \bibnamefont
  {Solodukhin}},\ }\bibfield  {title} {\bibinfo {title} {Entanglement entropy
  of black holes},\ }\href {https://doi.org/10.12942/lrr-2011-8} {\bibfield
  {journal} {\bibinfo  {journal} {Living Reviews in Relativity}\ }\textbf
  {\bibinfo {volume} {14}},\ \bibinfo {pages} {8} (\bibinfo {year}
  {2011})}\BibitemShut {NoStop}%
\bibitem [{\citenamefont {Hartman}\ and\ \citenamefont
  {Maldacena}(2013)}]{Hartman2013}%
  \BibitemOpen
  \bibfield  {author} {\bibinfo {author} {\bibfnamefont {T.}~\bibnamefont
  {Hartman}}\ and\ \bibinfo {author} {\bibfnamefont {J.}~\bibnamefont
  {Maldacena}},\ }\bibfield  {title} {\bibinfo {title} {Time evolution of
  entanglement entropy from black hole interiors},\ }\href
  {https://doi.org/10.1007/JHEP05(2013)014} {\bibfield  {journal} {\bibinfo
  {journal} {Journal of High Energy Physics}\ }\textbf {\bibinfo {volume}
  {2013}},\ \bibinfo {pages} {14} (\bibinfo {year} {2013})}\BibitemShut
  {NoStop}%
\bibitem [{\citenamefont {Bauer}\ and\ \citenamefont
  {Nayak}(2013)}]{Bauer_2013}%
  \BibitemOpen
  \bibfield  {author} {\bibinfo {author} {\bibfnamefont {B.}~\bibnamefont
  {Bauer}}\ and\ \bibinfo {author} {\bibfnamefont {C.}~\bibnamefont {Nayak}},\
  }\bibfield  {title} {\bibinfo {title} {Area laws in a many-body localized
  state and its implications for topological order},\ }\href
  {https://doi.org/10.1088/1742-5468/2013/09/P09005} {\bibfield  {journal}
  {\bibinfo  {journal} {Journal of Statistical Mechanics: Theory and
  Experiment}\ }\textbf {\bibinfo {volume} {2013}},\ \bibinfo {pages} {P09005}
  (\bibinfo {year} {2013})}\BibitemShut {NoStop}%
\bibitem [{\citenamefont {Leung}\ \emph {et~al.}(2017)\citenamefont {Leung},
  \citenamefont {Abdelhafez}, \citenamefont {Koch},\ and\ \citenamefont
  {Schuster}}]{PhysRevA.95.042318}%
  \BibitemOpen
  \bibfield  {author} {\bibinfo {author} {\bibfnamefont {N.}~\bibnamefont
  {Leung}}, \bibinfo {author} {\bibfnamefont {M.}~\bibnamefont {Abdelhafez}},
  \bibinfo {author} {\bibfnamefont {J.}~\bibnamefont {Koch}},\ and\ \bibinfo
  {author} {\bibfnamefont {D.}~\bibnamefont {Schuster}},\ }\bibfield  {title}
  {\bibinfo {title} {Speedup for quantum optimal control from automatic
  differentiation based on graphics processing units},\ }\href
  {https://doi.org/10.1103/PhysRevA.95.042318} {\bibfield  {journal} {\bibinfo
  {journal} {Phys. Rev. A}\ }\textbf {\bibinfo {volume} {95}},\ \bibinfo
  {pages} {042318} (\bibinfo {year} {2017})}\BibitemShut {NoStop}%
\bibitem [{\citenamefont {Coe}\ \emph {et~al.}(2009)\citenamefont {Coe},
  \citenamefont {Capelle},\ and\ \citenamefont {D'Amico}}]{PhysRevA.79.032504}%
  \BibitemOpen
  \bibfield  {author} {\bibinfo {author} {\bibfnamefont {J.~P.}\ \bibnamefont
  {Coe}}, \bibinfo {author} {\bibfnamefont {K.}~\bibnamefont {Capelle}},\ and\
  \bibinfo {author} {\bibfnamefont {I.}~\bibnamefont {D'Amico}},\ }\bibfield
  {title} {\bibinfo {title} {Reverse engineering in many-body quantum physics:
  Correspondence between many-body systems and effective single-particle
  equations},\ }\href {https://doi.org/10.1103/PhysRevA.79.032504} {\bibfield
  {journal} {\bibinfo  {journal} {Phys. Rev. A}\ }\textbf {\bibinfo {volume}
  {79}},\ \bibinfo {pages} {032504} (\bibinfo {year} {2009})}\BibitemShut
  {NoStop}%
\bibitem [{\citenamefont {Chertkov}\ and\ \citenamefont
  {Clark}(2018)}]{PhysRevX.8.031029}%
  \BibitemOpen
  \bibfield  {author} {\bibinfo {author} {\bibfnamefont {E.}~\bibnamefont
  {Chertkov}}\ and\ \bibinfo {author} {\bibfnamefont {B.~K.}\ \bibnamefont
  {Clark}},\ }\bibfield  {title} {\bibinfo {title} {Computational inverse
  method for constructing spaces of quantum models from wave functions},\
  }\href {https://doi.org/10.1103/PhysRevX.8.031029} {\bibfield  {journal}
  {\bibinfo  {journal} {Phys. Rev. X}\ }\textbf {\bibinfo {volume} {8}},\
  \bibinfo {pages} {031029} (\bibinfo {year} {2018})}\BibitemShut {NoStop}%
\bibitem [{\citenamefont {Pakrouski}(2020)}]{Pakrouski2020automaticdesignof}%
  \BibitemOpen
  \bibfield  {author} {\bibinfo {author} {\bibfnamefont {K.}~\bibnamefont
  {Pakrouski}},\ }\bibfield  {title} {\bibinfo {title} {Automatic design of
  {H}amiltonians},\ }\href {https://doi.org/10.22331/q-2020-09-02-315}
  {\bibfield  {journal} {\bibinfo  {journal} {{Quantum}}\ }\textbf {\bibinfo
  {volume} {4}},\ \bibinfo {pages} {315} (\bibinfo {year} {2020})}\BibitemShut
  {NoStop}%
\bibitem [{\citenamefont {Qi}\ and\ \citenamefont
  {Ranard}(2019)}]{Qi2019determininglocal}%
  \BibitemOpen
  \bibfield  {author} {\bibinfo {author} {\bibfnamefont {X.-L.}\ \bibnamefont
  {Qi}}\ and\ \bibinfo {author} {\bibfnamefont {D.}~\bibnamefont {Ranard}},\
  }\bibfield  {title} {\bibinfo {title} {Determining a local {H}amiltonian from
  a single eigenstate},\ }\href {https://doi.org/10.22331/q-2019-07-08-159}
  {\bibfield  {journal} {\bibinfo  {journal} {{Quantum}}\ }\textbf {\bibinfo
  {volume} {3}},\ \bibinfo {pages} {159} (\bibinfo {year} {2019})}\BibitemShut
  {NoStop}%
\bibitem [{\citenamefont {Greiter}\ \emph {et~al.}(2018)\citenamefont
  {Greiter}, \citenamefont {Schnells},\ and\ \citenamefont
  {Thomale}}]{PhysRevB.98.081113}%
  \BibitemOpen
  \bibfield  {author} {\bibinfo {author} {\bibfnamefont {M.}~\bibnamefont
  {Greiter}}, \bibinfo {author} {\bibfnamefont {V.}~\bibnamefont {Schnells}},\
  and\ \bibinfo {author} {\bibfnamefont {R.}~\bibnamefont {Thomale}},\
  }\bibfield  {title} {\bibinfo {title} {Method to identify parent hamiltonians
  for trial states},\ }\href {https://doi.org/10.1103/PhysRevB.98.081113}
  {\bibfield  {journal} {\bibinfo  {journal} {Phys. Rev. B}\ }\textbf {\bibinfo
  {volume} {98}},\ \bibinfo {pages} {081113} (\bibinfo {year}
  {2018})}\BibitemShut {NoStop}%
\bibitem [{\citenamefont {Fujita}\ \emph {et~al.}(2018)\citenamefont {Fujita},
  \citenamefont {Nakagawa}, \citenamefont {Sugiura},\ and\ \citenamefont
  {Oshikawa}}]{PhysRevB.97.075114}%
  \BibitemOpen
  \bibfield  {author} {\bibinfo {author} {\bibfnamefont {H.}~\bibnamefont
  {Fujita}}, \bibinfo {author} {\bibfnamefont {Y.~O.}\ \bibnamefont
  {Nakagawa}}, \bibinfo {author} {\bibfnamefont {S.}~\bibnamefont {Sugiura}},\
  and\ \bibinfo {author} {\bibfnamefont {M.}~\bibnamefont {Oshikawa}},\
  }\bibfield  {title} {\bibinfo {title} {Construction of hamiltonians by
  supervised learning of energy and entanglement spectra},\ }\href
  {https://doi.org/10.1103/PhysRevB.97.075114} {\bibfield  {journal} {\bibinfo
  {journal} {Phys. Rev. B}\ }\textbf {\bibinfo {volume} {97}},\ \bibinfo
  {pages} {075114} (\bibinfo {year} {2018})}\BibitemShut {NoStop}%
\bibitem [{\citenamefont {Tang}\ and\ \citenamefont
  {Zhu}(2023)}]{tang2023nonhermitian}%
  \BibitemOpen
  \bibfield  {author} {\bibinfo {author} {\bibfnamefont {Y.}~\bibnamefont
  {Tang}}\ and\ \bibinfo {author} {\bibfnamefont {W.}~\bibnamefont {Zhu}},\
  }\href@noop {} {\bibinfo {title} {Non-hermitian parent hamiltonian from
  generalized quantum covariance matrix}} (\bibinfo {year} {2023}),\ \Eprint
  {https://arxiv.org/abs/2307.03107} {arXiv:2307.03107 [quant-ph]} \BibitemShut
  {NoStop}%
\bibitem [{\citenamefont {Tamura}\ and\ \citenamefont
  {Hukushima}(2017)}]{PhysRevB.95.064407}%
  \BibitemOpen
  \bibfield  {author} {\bibinfo {author} {\bibfnamefont {R.}~\bibnamefont
  {Tamura}}\ and\ \bibinfo {author} {\bibfnamefont {K.}~\bibnamefont
  {Hukushima}},\ }\bibfield  {title} {\bibinfo {title} {Method for estimating
  spin-spin interactions from magnetization curves},\ }\href
  {https://doi.org/10.1103/PhysRevB.95.064407} {\bibfield  {journal} {\bibinfo
  {journal} {Phys. Rev. B}\ }\textbf {\bibinfo {volume} {95}},\ \bibinfo
  {pages} {064407} (\bibinfo {year} {2017})}\BibitemShut {NoStop}%
\bibitem [{\citenamefont {Sanchez-Lengeling}\ and\ \citenamefont
  {Aspuru-Guzik}(2018)}]{doi:10.1126/science.aat2663}%
  \BibitemOpen
  \bibfield  {author} {\bibinfo {author} {\bibfnamefont {B.}~\bibnamefont
  {Sanchez-Lengeling}}\ and\ \bibinfo {author} {\bibfnamefont {A.}~\bibnamefont
  {Aspuru-Guzik}},\ }\bibfield  {title} {\bibinfo {title} {Inverse molecular
  design using machine learning: Generative models for matter engineering},\
  }\href {https://doi.org/10.1126/science.aat2663} {\bibfield  {journal}
  {\bibinfo  {journal} {Science}\ }\textbf {\bibinfo {volume} {361}},\ \bibinfo
  {pages} {360} (\bibinfo {year} {2018})}\BibitemShut {NoStop}%
\bibitem [{\citenamefont {Mertz}\ and\ \citenamefont
  {Valent\'{\i}}(2021)}]{PhysRevResearch.3.013132}%
  \BibitemOpen
  \bibfield  {author} {\bibinfo {author} {\bibfnamefont {T.}~\bibnamefont
  {Mertz}}\ and\ \bibinfo {author} {\bibfnamefont {R.}~\bibnamefont
  {Valent\'{\i}}},\ }\bibfield  {title} {\bibinfo {title} {Engineering
  topological phases guided by statistical and machine learning methods},\
  }\href {https://doi.org/10.1103/PhysRevResearch.3.013132} {\bibfield
  {journal} {\bibinfo  {journal} {Phys. Rev. Res.}\ }\textbf {\bibinfo {volume}
  {3}},\ \bibinfo {pages} {013132} (\bibinfo {year} {2021})}\BibitemShut
  {NoStop}%
\bibitem [{\citenamefont {Inui}\ and\ \citenamefont {Motome}(2023)}]{Inui2023}%
  \BibitemOpen
  \bibfield  {author} {\bibinfo {author} {\bibfnamefont {K.}~\bibnamefont
  {Inui}}\ and\ \bibinfo {author} {\bibfnamefont {Y.}~\bibnamefont {Motome}},\
  }\bibfield  {title} {\bibinfo {title} {Inverse hamiltonian design by
  automatic differentiation},\ }\href
  {https://doi.org/10.1038/s42005-023-01132-0} {\bibfield  {journal} {\bibinfo
  {journal} {Communications Physics}\ }\textbf {\bibinfo {volume} {6}},\
  \bibinfo {pages} {37} (\bibinfo {year} {2023})}\BibitemShut {NoStop}%
\bibitem [{\citenamefont {Plenio}(2005)}]{PhysRevLett.95.090503}%
  \BibitemOpen
  \bibfield  {author} {\bibinfo {author} {\bibfnamefont {M.~B.}\ \bibnamefont
  {Plenio}},\ }\bibfield  {title} {\bibinfo {title} {Logarithmic negativity: A
  full entanglement monotone that is not convex},\ }\href
  {https://doi.org/10.1103/PhysRevLett.95.090503} {\bibfield  {journal}
  {\bibinfo  {journal} {Phys. Rev. Lett.}\ }\textbf {\bibinfo {volume} {95}},\
  \bibinfo {pages} {090503} (\bibinfo {year} {2005})}\BibitemShut {NoStop}%
\bibitem [{\citenamefont {Vedral}\ \emph {et~al.}(1997)\citenamefont {Vedral},
  \citenamefont {Plenio}, \citenamefont {Rippin},\ and\ \citenamefont
  {Knight}}]{PhysRevLett.78.2275}%
  \BibitemOpen
  \bibfield  {author} {\bibinfo {author} {\bibfnamefont {V.}~\bibnamefont
  {Vedral}}, \bibinfo {author} {\bibfnamefont {M.~B.}\ \bibnamefont {Plenio}},
  \bibinfo {author} {\bibfnamefont {M.~A.}\ \bibnamefont {Rippin}},\ and\
  \bibinfo {author} {\bibfnamefont {P.~L.}\ \bibnamefont {Knight}},\ }\bibfield
   {title} {\bibinfo {title} {Quantifying entanglement},\ }\href
  {https://doi.org/10.1103/PhysRevLett.78.2275} {\bibfield  {journal} {\bibinfo
   {journal} {Phys. Rev. Lett.}\ }\textbf {\bibinfo {volume} {78}},\ \bibinfo
  {pages} {2275} (\bibinfo {year} {1997})}\BibitemShut {NoStop}%
\bibitem [{\citenamefont {Kitaev}(2006)}]{KITAEV20062}%
  \BibitemOpen
  \bibfield  {author} {\bibinfo {author} {\bibfnamefont {A.}~\bibnamefont
  {Kitaev}},\ }\bibfield  {title} {\bibinfo {title} {Anyons in an exactly
  solved model and beyond},\ }\href
  {https://doi.org/https://doi.org/10.1016/j.aop.2005.10.005} {\bibfield
  {journal} {\bibinfo  {journal} {Annals of Physics}\ }\textbf {\bibinfo
  {volume} {321}},\ \bibinfo {pages} {2} (\bibinfo {year} {2006})},\ \bibinfo
  {note} {january Special Issue}\BibitemShut {NoStop}%
\bibitem [{\citenamefont {Anderson}(1973)}]{ANDERSON1973153}%
  \BibitemOpen
  \bibfield  {author} {\bibinfo {author} {\bibfnamefont {P.}~\bibnamefont
  {Anderson}},\ }\bibfield  {title} {\bibinfo {title} {Resonating valence
  bonds: A new kind of insulator?},\ }\href
  {https://doi.org/https://doi.org/10.1016/0025-5408(73)90167-0} {\bibfield
  {journal} {\bibinfo  {journal} {Materials Research Bulletin}\ }\textbf
  {\bibinfo {volume} {8}},\ \bibinfo {pages} {153} (\bibinfo {year}
  {1973})}\BibitemShut {NoStop}%
\bibitem [{\citenamefont {Bennett}\ \emph
  {et~al.}(1996{\natexlab{b}})\citenamefont {Bennett}, \citenamefont
  {Bernstein}, \citenamefont {Popescu},\ and\ \citenamefont
  {Schumacher}}]{PhysRevA.53.2046}%
  \BibitemOpen
  \bibfield  {author} {\bibinfo {author} {\bibfnamefont {C.~H.}\ \bibnamefont
  {Bennett}}, \bibinfo {author} {\bibfnamefont {H.~J.}\ \bibnamefont
  {Bernstein}}, \bibinfo {author} {\bibfnamefont {S.}~\bibnamefont {Popescu}},\
  and\ \bibinfo {author} {\bibfnamefont {B.}~\bibnamefont {Schumacher}},\
  }\bibfield  {title} {\bibinfo {title} {Concentrating partial entanglement by
  local operations},\ }\href {https://doi.org/10.1103/PhysRevA.53.2046}
  {\bibfield  {journal} {\bibinfo  {journal} {Phys. Rev. A}\ }\textbf {\bibinfo
  {volume} {53}},\ \bibinfo {pages} {2046} (\bibinfo {year}
  {1996}{\natexlab{b}})}\BibitemShut {NoStop}%
\bibitem [{\citenamefont {Zhuang}\ \emph {et~al.}(2020)\citenamefont {Zhuang},
  \citenamefont {Tang}, \citenamefont {Ding}, \citenamefont {Tatikonda},
  \citenamefont {Dvornek}, \citenamefont {Papademetris},\ and\ \citenamefont
  {Duncan}}]{NEURIPS2020_d9d4f495}%
  \BibitemOpen
  \bibfield  {author} {\bibinfo {author} {\bibfnamefont {J.}~\bibnamefont
  {Zhuang}}, \bibinfo {author} {\bibfnamefont {T.}~\bibnamefont {Tang}},
  \bibinfo {author} {\bibfnamefont {Y.}~\bibnamefont {Ding}}, \bibinfo {author}
  {\bibfnamefont {S.~C.}\ \bibnamefont {Tatikonda}}, \bibinfo {author}
  {\bibfnamefont {N.}~\bibnamefont {Dvornek}}, \bibinfo {author} {\bibfnamefont
  {X.}~\bibnamefont {Papademetris}},\ and\ \bibinfo {author} {\bibfnamefont
  {J.}~\bibnamefont {Duncan}},\ }\bibfield  {title} {\bibinfo {title}
  {Adabelief optimizer: Adapting stepsizes by the belief in observed
  gradients},\ }in\ \href
  {https://proceedings.neurips.cc/paper_files/paper/2020/file/d9d4f495e875a2e075a1a4a6e1b9770f-Paper.pdf}
  {\emph {\bibinfo {booktitle} {Advances in Neural Information Processing
  Systems}}},\ Vol.~\bibinfo {volume} {33},\ \bibinfo {editor} {edited by\
  \bibinfo {editor} {\bibfnamefont {H.}~\bibnamefont {Larochelle}}, \bibinfo
  {editor} {\bibfnamefont {M.}~\bibnamefont {Ranzato}}, \bibinfo {editor}
  {\bibfnamefont {R.}~\bibnamefont {Hadsell}}, \bibinfo {editor} {\bibfnamefont
  {M.}~\bibnamefont {Balcan}},\ and\ \bibinfo {editor} {\bibfnamefont
  {H.}~\bibnamefont {Lin}}}\ (\bibinfo  {publisher} {Curran Associates, Inc.},\
  \bibinfo {year} {2020})\ pp.\ \bibinfo {pages} {18795--18806}\BibitemShut
  {NoStop}%
\bibitem [{\citenamefont {Bradbury}\ \emph {et~al.}(2018)\citenamefont
  {Bradbury}, \citenamefont {Frostig}, \citenamefont {Hawkins}, \citenamefont
  {Johnson}, \citenamefont {Leary}, \citenamefont {Maclaurin}, \citenamefont
  {Necula}, \citenamefont {Paszke}, \citenamefont {Vander{P}las}, \citenamefont
  {Wanderman-{M}ilne},\ and\ \citenamefont {Zhang}}]{jax2018github}%
  \BibitemOpen
  \bibfield  {author} {\bibinfo {author} {\bibfnamefont {J.}~\bibnamefont
  {Bradbury}}, \bibinfo {author} {\bibfnamefont {R.}~\bibnamefont {Frostig}},
  \bibinfo {author} {\bibfnamefont {P.}~\bibnamefont {Hawkins}}, \bibinfo
  {author} {\bibfnamefont {M.~J.}\ \bibnamefont {Johnson}}, \bibinfo {author}
  {\bibfnamefont {C.}~\bibnamefont {Leary}}, \bibinfo {author} {\bibfnamefont
  {D.}~\bibnamefont {Maclaurin}}, \bibinfo {author} {\bibfnamefont
  {G.}~\bibnamefont {Necula}}, \bibinfo {author} {\bibfnamefont
  {A.}~\bibnamefont {Paszke}}, \bibinfo {author} {\bibfnamefont
  {J.}~\bibnamefont {Vander{P}las}}, \bibinfo {author} {\bibfnamefont
  {S.}~\bibnamefont {Wanderman-{M}ilne}},\ and\ \bibinfo {author}
  {\bibfnamefont {Q.}~\bibnamefont {Zhang}},\ }\href
  {http://github.com/google/jax} {\bibinfo {title} {{JAX}: composable
  transformations of {P}ython+{N}um{P}y programs}} (\bibinfo {year}
  {2018})\BibitemShut {NoStop}%
\bibitem [{\citenamefont {Nussinov}\ and\ \citenamefont {van~den
  Brink}(2015)}]{RevModPhys.87.1}%
  \BibitemOpen
  \bibfield  {author} {\bibinfo {author} {\bibfnamefont {Z.}~\bibnamefont
  {Nussinov}}\ and\ \bibinfo {author} {\bibfnamefont {J.}~\bibnamefont {van~den
  Brink}},\ }\bibfield  {title} {\bibinfo {title} {Compass models: Theory and
  physical motivations},\ }\href {https://doi.org/10.1103/RevModPhys.87.1}
  {\bibfield  {journal} {\bibinfo  {journal} {Rev. Mod. Phys.}\ }\textbf
  {\bibinfo {volume} {87}},\ \bibinfo {pages} {1} (\bibinfo {year}
  {2015})}\BibitemShut {NoStop}%
\bibitem [{\citenamefont {Taniguchi}\ \emph {et~al.}(1995)\citenamefont
  {Taniguchi}, \citenamefont {Nishikawa}, \citenamefont {Yasui}, \citenamefont
  {Kobayashi}, \citenamefont {Sato}, \citenamefont {Nishioka}, \citenamefont
  {Kontani},\ and\ \citenamefont {Sano}}]{doi:10.1143/JPSJ.64.2758}%
  \BibitemOpen
  \bibfield  {author} {\bibinfo {author} {\bibfnamefont {S.}~\bibnamefont
  {Taniguchi}}, \bibinfo {author} {\bibfnamefont {T.}~\bibnamefont
  {Nishikawa}}, \bibinfo {author} {\bibfnamefont {Y.}~\bibnamefont {Yasui}},
  \bibinfo {author} {\bibfnamefont {Y.}~\bibnamefont {Kobayashi}}, \bibinfo
  {author} {\bibfnamefont {M.}~\bibnamefont {Sato}}, \bibinfo {author}
  {\bibfnamefont {T.}~\bibnamefont {Nishioka}}, \bibinfo {author}
  {\bibfnamefont {M.}~\bibnamefont {Kontani}},\ and\ \bibinfo {author}
  {\bibfnamefont {K.}~\bibnamefont {Sano}},\ }\bibfield  {title} {\bibinfo
  {title} {Spin gap behavior of s=1/2 quasi-two-dimensional system cav4o9},\
  }\href {https://doi.org/10.1143/JPSJ.64.2758} {\bibfield  {journal} {\bibinfo
   {journal} {Journal of the Physical Society of Japan}\ }\textbf {\bibinfo
  {volume} {64}},\ \bibinfo {pages} {2758} (\bibinfo {year}
  {1995})}\BibitemShut {NoStop}%
\bibitem [{\citenamefont {Landau}\ \emph {et~al.}(2016)\citenamefont {Landau},
  \citenamefont {Plugge}, \citenamefont {Sela}, \citenamefont {Altland},
  \citenamefont {Albrecht},\ and\ \citenamefont
  {Egger}}]{PhysRevLett.116.050501}%
  \BibitemOpen
  \bibfield  {author} {\bibinfo {author} {\bibfnamefont {L.~A.}\ \bibnamefont
  {Landau}}, \bibinfo {author} {\bibfnamefont {S.}~\bibnamefont {Plugge}},
  \bibinfo {author} {\bibfnamefont {E.}~\bibnamefont {Sela}}, \bibinfo {author}
  {\bibfnamefont {A.}~\bibnamefont {Altland}}, \bibinfo {author} {\bibfnamefont
  {S.~M.}\ \bibnamefont {Albrecht}},\ and\ \bibinfo {author} {\bibfnamefont
  {R.}~\bibnamefont {Egger}},\ }\bibfield  {title} {\bibinfo {title} {Towards
  realistic implementations of a majorana surface code},\ }\href
  {https://doi.org/10.1103/PhysRevLett.116.050501} {\bibfield  {journal}
  {\bibinfo  {journal} {Phys. Rev. Lett.}\ }\textbf {\bibinfo {volume} {116}},\
  \bibinfo {pages} {050501} (\bibinfo {year} {2016})}\BibitemShut {NoStop}%
\bibitem [{\citenamefont {Litinski}\ and\ \citenamefont {von
  Oppen}(2018)}]{PhysRevB.97.205404}%
  \BibitemOpen
  \bibfield  {author} {\bibinfo {author} {\bibfnamefont {D.}~\bibnamefont
  {Litinski}}\ and\ \bibinfo {author} {\bibfnamefont {F.}~\bibnamefont {von
  Oppen}},\ }\bibfield  {title} {\bibinfo {title} {Quantum computing with
  majorana fermion codes},\ }\href {https://doi.org/10.1103/PhysRevB.97.205404}
  {\bibfield  {journal} {\bibinfo  {journal} {Phys. Rev. B}\ }\textbf {\bibinfo
  {volume} {97}},\ \bibinfo {pages} {205404} (\bibinfo {year}
  {2018})}\BibitemShut {NoStop}%
\bibitem [{\citenamefont {Yang}\ \emph {et~al.}(2007)\citenamefont {Yang},
  \citenamefont {Zhou},\ and\ \citenamefont {Sun}}]{PhysRevB.76.180404}%
  \BibitemOpen
  \bibfield  {author} {\bibinfo {author} {\bibfnamefont {S.}~\bibnamefont
  {Yang}}, \bibinfo {author} {\bibfnamefont {D.~L.}\ \bibnamefont {Zhou}},\
  and\ \bibinfo {author} {\bibfnamefont {C.~P.}\ \bibnamefont {Sun}},\
  }\bibfield  {title} {\bibinfo {title} {Mosaic spin models with topological
  order},\ }\href {https://doi.org/10.1103/PhysRevB.76.180404} {\bibfield
  {journal} {\bibinfo  {journal} {Phys. Rev. B}\ }\textbf {\bibinfo {volume}
  {76}},\ \bibinfo {pages} {180404} (\bibinfo {year} {2007})}\BibitemShut
  {NoStop}%
\bibitem [{\citenamefont {Kells}\ \emph {et~al.}(2011)\citenamefont {Kells},
  \citenamefont {Kailasvuori}, \citenamefont {Slingerland},\ and\ \citenamefont
  {Vala}}]{Kells_2011}%
  \BibitemOpen
  \bibfield  {author} {\bibinfo {author} {\bibfnamefont {G.}~\bibnamefont
  {Kells}}, \bibinfo {author} {\bibfnamefont {J.}~\bibnamefont {Kailasvuori}},
  \bibinfo {author} {\bibfnamefont {J.~K.}\ \bibnamefont {Slingerland}},\ and\
  \bibinfo {author} {\bibfnamefont {J.}~\bibnamefont {Vala}},\ }\bibfield
  {title} {\bibinfo {title} {Kaleidoscope of topological phases with multiple
  majorana species},\ }\href {https://doi.org/10.1088/1367-2630/13/9/095014}
  {\bibfield  {journal} {\bibinfo  {journal} {New Journal of Physics}\ }\textbf
  {\bibinfo {volume} {13}},\ \bibinfo {pages} {095014} (\bibinfo {year}
  {2011})}\BibitemShut {NoStop}%
\bibitem [{\citenamefont {Balents}(2010)}]{Balents2010}%
  \BibitemOpen
  \bibfield  {author} {\bibinfo {author} {\bibfnamefont {L.}~\bibnamefont
  {Balents}},\ }\bibfield  {title} {\bibinfo {title} {Spin liquids in
  frustrated magnets},\ }\href {https://doi.org/10.1038/nature08917} {\bibfield
   {journal} {\bibinfo  {journal} {Nature}\ }\textbf {\bibinfo {volume}
  {464}},\ \bibinfo {pages} {199} (\bibinfo {year} {2010})}\BibitemShut
  {NoStop}%
\bibitem [{\citenamefont {Bernu}\ \emph {et~al.}(1994)\citenamefont {Bernu},
  \citenamefont {Lecheminant}, \citenamefont {Lhuillier},\ and\ \citenamefont
  {Pierre}}]{PhysRevB.50.10048}%
  \BibitemOpen
  \bibfield  {author} {\bibinfo {author} {\bibfnamefont {B.}~\bibnamefont
  {Bernu}}, \bibinfo {author} {\bibfnamefont {P.}~\bibnamefont {Lecheminant}},
  \bibinfo {author} {\bibfnamefont {C.}~\bibnamefont {Lhuillier}},\ and\
  \bibinfo {author} {\bibfnamefont {L.}~\bibnamefont {Pierre}},\ }\bibfield
  {title} {\bibinfo {title} {Exact spectra, spin susceptibilities, and order
  parameter of the quantum heisenberg antiferromagnet on the triangular
  lattice},\ }\href {https://doi.org/10.1103/PhysRevB.50.10048} {\bibfield
  {journal} {\bibinfo  {journal} {Phys. Rev. B}\ }\textbf {\bibinfo {volume}
  {50}},\ \bibinfo {pages} {10048} (\bibinfo {year} {1994})}\BibitemShut
  {NoStop}%
\bibitem [{\citenamefont {Capriotti}\ \emph {et~al.}(1999)\citenamefont
  {Capriotti}, \citenamefont {Trumper},\ and\ \citenamefont
  {Sorella}}]{PhysRevLett.82.3899}%
  \BibitemOpen
  \bibfield  {author} {\bibinfo {author} {\bibfnamefont {L.}~\bibnamefont
  {Capriotti}}, \bibinfo {author} {\bibfnamefont {A.~E.}\ \bibnamefont
  {Trumper}},\ and\ \bibinfo {author} {\bibfnamefont {S.}~\bibnamefont
  {Sorella}},\ }\bibfield  {title} {\bibinfo {title} {Long-range n\'eel order
  in the triangular heisenberg model},\ }\href
  {https://doi.org/10.1103/PhysRevLett.82.3899} {\bibfield  {journal} {\bibinfo
   {journal} {Phys. Rev. Lett.}\ }\textbf {\bibinfo {volume} {82}},\ \bibinfo
  {pages} {3899} (\bibinfo {year} {1999})}\BibitemShut {NoStop}%
\bibitem [{\citenamefont {Zheng}\ \emph {et~al.}(2006)\citenamefont {Zheng},
  \citenamefont {Fj\ae{}restad}, \citenamefont {Singh}, \citenamefont
  {McKenzie},\ and\ \citenamefont {Coldea}}]{PhysRevB.74.224420}%
  \BibitemOpen
  \bibfield  {author} {\bibinfo {author} {\bibfnamefont {W.}~\bibnamefont
  {Zheng}}, \bibinfo {author} {\bibfnamefont {J.~O.}\ \bibnamefont
  {Fj\ae{}restad}}, \bibinfo {author} {\bibfnamefont {R.~R.~P.}\ \bibnamefont
  {Singh}}, \bibinfo {author} {\bibfnamefont {R.~H.}\ \bibnamefont
  {McKenzie}},\ and\ \bibinfo {author} {\bibfnamefont {R.}~\bibnamefont
  {Coldea}},\ }\bibfield  {title} {\bibinfo {title} {Excitation spectra of the
  spin-$\frac{1}{2}$ triangular-lattice heisenberg antiferromagnet},\ }\href
  {https://doi.org/10.1103/PhysRevB.74.224420} {\bibfield  {journal} {\bibinfo
  {journal} {Phys. Rev. B}\ }\textbf {\bibinfo {volume} {74}},\ \bibinfo
  {pages} {224420} (\bibinfo {year} {2006})}\BibitemShut {NoStop}%
\bibitem [{\citenamefont {Castro}\ \emph {et~al.}(2006)\citenamefont {Castro},
  \citenamefont {Peres}, \citenamefont {Beach},\ and\ \citenamefont
  {Sandvik}}]{PhysRevB.73.054422}%
  \BibitemOpen
  \bibfield  {author} {\bibinfo {author} {\bibfnamefont {E.~V.}\ \bibnamefont
  {Castro}}, \bibinfo {author} {\bibfnamefont {N.~M.~R.}\ \bibnamefont
  {Peres}}, \bibinfo {author} {\bibfnamefont {K.~S.~D.}\ \bibnamefont
  {Beach}},\ and\ \bibinfo {author} {\bibfnamefont {A.~W.}\ \bibnamefont
  {Sandvik}},\ }\bibfield  {title} {\bibinfo {title} {Site dilution of quantum
  spins in the honeycomb lattice},\ }\href
  {https://doi.org/10.1103/PhysRevB.73.054422} {\bibfield  {journal} {\bibinfo
  {journal} {Phys. Rev. B}\ }\textbf {\bibinfo {volume} {73}},\ \bibinfo
  {pages} {054422} (\bibinfo {year} {2006})}\BibitemShut {NoStop}%
\bibitem [{\citenamefont {Fennell}\ \emph {et~al.}(2011)\citenamefont
  {Fennell}, \citenamefont {Piatek}, \citenamefont {Stephenson}, \citenamefont
  {Nilsen},\ and\ \citenamefont {R^^c3^^b8nnow}}]{Fennell_2011}%
  \BibitemOpen
  \bibfield  {author} {\bibinfo {author} {\bibfnamefont {T.}~\bibnamefont
  {Fennell}}, \bibinfo {author} {\bibfnamefont {J.~O.}\ \bibnamefont {Piatek}},
  \bibinfo {author} {\bibfnamefont {R.~A.}\ \bibnamefont {Stephenson}},
  \bibinfo {author} {\bibfnamefont {G.~J.}\ \bibnamefont {Nilsen}},\ and\
  \bibinfo {author} {\bibfnamefont {H.~M.}\ \bibnamefont {R^^c3^^b8nnow}},\
  }\bibfield  {title} {\bibinfo {title} {Spangolite: an s = 1/2 maple leaf
  lattice antiferromagnet?},\ }\href
  {https://doi.org/10.1088/0953-8984/23/16/164201} {\bibfield  {journal}
  {\bibinfo  {journal} {Journal of Physics: Condensed Matter}\ }\textbf
  {\bibinfo {volume} {23}},\ \bibinfo {pages} {164201} (\bibinfo {year}
  {2011})}\BibitemShut {NoStop}%
\bibitem [{\citenamefont {Aliev}\ \emph {et~al.}(2012)\citenamefont {Aliev},
  \citenamefont {Huv^^c3^^a9}, \citenamefont {Colis}, \citenamefont {Colmont},
  \citenamefont {Dinia},\ and\ \citenamefont
  {Mentr^^c3^^a9}}]{https://doi.org/10.1002/anie.201203775}%
  \BibitemOpen
  \bibfield  {author} {\bibinfo {author} {\bibfnamefont {A.}~\bibnamefont
  {Aliev}}, \bibinfo {author} {\bibfnamefont {M.}~\bibnamefont {Huv^^c3^^a9}},
  \bibinfo {author} {\bibfnamefont {S.}~\bibnamefont {Colis}}, \bibinfo
  {author} {\bibfnamefont {M.}~\bibnamefont {Colmont}}, \bibinfo {author}
  {\bibfnamefont {A.}~\bibnamefont {Dinia}},\ and\ \bibinfo {author}
  {\bibfnamefont {O.}~\bibnamefont {Mentr^^c3^^a9}},\ }\bibfield  {title}
  {\bibinfo {title} {Two-dimensional antiferromagnetism in the
  [{M}n$_{3+x}${O}$_7$][{B}i$_4${O}$_{4.5 - y}$] compound with a maple-leaf
  lattice},\ }\href {https://doi.org/https://doi.org/10.1002/anie.201203775}
  {\bibfield  {journal} {\bibinfo  {journal} {Angewandte Chemie International
  Edition}\ }\textbf {\bibinfo {volume} {51}},\ \bibinfo {pages} {9393}
  (\bibinfo {year} {2012})}\BibitemShut {NoStop}%
\bibitem [{\citenamefont {Yao}\ and\ \citenamefont
  {Qi}(2010)}]{PhysRevLett.105.080501}%
  \BibitemOpen
  \bibfield  {author} {\bibinfo {author} {\bibfnamefont {H.}~\bibnamefont
  {Yao}}\ and\ \bibinfo {author} {\bibfnamefont {X.-L.}\ \bibnamefont {Qi}},\
  }\bibfield  {title} {\bibinfo {title} {Entanglement entropy and entanglement
  spectrum of the kitaev model},\ }\href
  {https://doi.org/10.1103/PhysRevLett.105.080501} {\bibfield  {journal}
  {\bibinfo  {journal} {Phys. Rev. Lett.}\ }\textbf {\bibinfo {volume} {105}},\
  \bibinfo {pages} {080501} (\bibinfo {year} {2010})}\BibitemShut {NoStop}%
\bibitem [{\citenamefont {Liao}\ \emph {et~al.}(2019)\citenamefont {Liao},
  \citenamefont {Liu}, \citenamefont {Wang},\ and\ \citenamefont
  {Xiang}}]{PhysRevX.9.031041}%
  \BibitemOpen
  \bibfield  {author} {\bibinfo {author} {\bibfnamefont {H.-J.}\ \bibnamefont
  {Liao}}, \bibinfo {author} {\bibfnamefont {J.-G.}\ \bibnamefont {Liu}},
  \bibinfo {author} {\bibfnamefont {L.}~\bibnamefont {Wang}},\ and\ \bibinfo
  {author} {\bibfnamefont {T.}~\bibnamefont {Xiang}},\ }\bibfield  {title}
  {\bibinfo {title} {Differentiable programming tensor networks},\ }\href
  {https://doi.org/10.1103/PhysRevX.9.031041} {\bibfield  {journal} {\bibinfo
  {journal} {Phys. Rev. X}\ }\textbf {\bibinfo {volume} {9}},\ \bibinfo {pages}
  {031041} (\bibinfo {year} {2019})}\BibitemShut {NoStop}%
\bibitem [{\citenamefont {Larkin}\ and\ \citenamefont
  {Ovchinnikov}(1969)}]{larkin1969quasiclassical}%
  \BibitemOpen
  \bibfield  {author} {\bibinfo {author} {\bibfnamefont {A.~I.}\ \bibnamefont
  {Larkin}}\ and\ \bibinfo {author} {\bibfnamefont {Y.~N.}\ \bibnamefont
  {Ovchinnikov}},\ }\bibfield  {title} {\bibinfo {title} {Quasiclassical method
  in the theory of superconductivity},\ }\href@noop {} {\bibfield  {journal}
  {\bibinfo  {journal} {Sov Phys JETP}\ }\textbf {\bibinfo {volume} {28}},\
  \bibinfo {pages} {1200} (\bibinfo {year} {1969})}\BibitemShut {NoStop}%
\bibitem [{\citenamefont {Shenker}\ and\ \citenamefont
  {Stanford}(2014)}]{Shenker2014}%
  \BibitemOpen
  \bibfield  {author} {\bibinfo {author} {\bibfnamefont {S.~H.}\ \bibnamefont
  {Shenker}}\ and\ \bibinfo {author} {\bibfnamefont {D.}~\bibnamefont
  {Stanford}},\ }\bibfield  {title} {\bibinfo {title} {Black holes and the
  butterfly effect},\ }\href {https://doi.org/10.1007/JHEP03(2014)067}
  {\bibfield  {journal} {\bibinfo  {journal} {Journal of High Energy Physics}\
  }\textbf {\bibinfo {volume} {2014}},\ \bibinfo {pages} {67} (\bibinfo {year}
  {2014})}\BibitemShut {NoStop}%
\bibitem [{\citenamefont {Maldacena}\ \emph {et~al.}(2016)\citenamefont
  {Maldacena}, \citenamefont {Shenker},\ and\ \citenamefont
  {Stanford}}]{Maldacena2016}%
  \BibitemOpen
  \bibfield  {author} {\bibinfo {author} {\bibfnamefont {J.}~\bibnamefont
  {Maldacena}}, \bibinfo {author} {\bibfnamefont {S.~H.}\ \bibnamefont
  {Shenker}},\ and\ \bibinfo {author} {\bibfnamefont {D.}~\bibnamefont
  {Stanford}},\ }\bibfield  {title} {\bibinfo {title} {A bound on chaos},\
  }\href {https://doi.org/10.1007/JHEP08(2016)106} {\bibfield  {journal}
  {\bibinfo  {journal} {Journal of High Energy Physics}\ }\textbf {\bibinfo
  {volume} {2016}},\ \bibinfo {pages} {106} (\bibinfo {year}
  {2016})}\BibitemShut {NoStop}%
\bibitem [{\citenamefont {Iyoda}\ and\ \citenamefont
  {Sagawa}(2018)}]{PhysRevA.97.042330}%
  \BibitemOpen
  \bibfield  {author} {\bibinfo {author} {\bibfnamefont {E.}~\bibnamefont
  {Iyoda}}\ and\ \bibinfo {author} {\bibfnamefont {T.}~\bibnamefont {Sagawa}},\
  }\bibfield  {title} {\bibinfo {title} {Scrambling of quantum information in
  quantum many-body systems},\ }\href
  {https://doi.org/10.1103/PhysRevA.97.042330} {\bibfield  {journal} {\bibinfo
  {journal} {Phys. Rev. A}\ }\textbf {\bibinfo {volume} {97}},\ \bibinfo
  {pages} {042330} (\bibinfo {year} {2018})}\BibitemShut {NoStop}%
\bibitem [{\citenamefont {Vidal}(2000)}]{doi:10.1080/09500340008244048}%
  \BibitemOpen
  \bibfield  {author} {\bibinfo {author} {\bibfnamefont {G.}~\bibnamefont
  {Vidal}},\ }\bibfield  {title} {\bibinfo {title} {Entanglement monotones},\
  }\href {https://doi.org/10.1080/09500340008244048} {\bibfield  {journal}
  {\bibinfo  {journal} {Journal of Modern Optics}\ }\textbf {\bibinfo {volume}
  {47}},\ \bibinfo {pages} {355} (\bibinfo {year} {2000})}\BibitemShut
  {NoStop}%
\end{thebibliography}%

\end{document}